# Magnons in Ferromagnetic Metallic Manganites

Jiandi Zhang[1,*], F. Ye[2], Hao Sha[1], Pengcheng Dai[3,2], J. A. Fernandez-Baca[2], and E.W. Plummer[3,4]

[1]*Department of physics, Florida International University, Miami, FL 33199, USA*

[2]*Center for Neutron Scattering, Oak Ridge National Laboratory, Oak Ridge, TN 37831, USA*

[3]*Department of physics & Astronomy, the University of Tennessee, Knoxville, TN 37996, USA*

[4]*Materials Science and Technology Division, Oak Ridge National Laboratory, Oak Ridge, TN 37831, USA*

* Electronic address: zhangj@fiu.edu

## ABSTRACT

**Ferromagnetic (FM) manganites, a group of likely half-metallic oxides, are of special interest not only because they are a testing ground of the classical double-exchange interaction mechanism for the "colossal" magnetoresistance, but also because they exhibit an extraordinary arena of emergent phenomena. These emergent phenomena are related to the complexity associated with strong interplay between charge, spin, orbital, and lattice. In this review, we focus on the use of inelastic neutron scattering to study the spin dynamics, mainly the magnon excitations in this class of FM metallic materials. In particular, we discussed the unusual magnon softening and damping near the Brillouin zone boundary in relatively narrow band compounds with strong Jahn-Teller lattice distortion and charge/orbital correlations. The anomalous behaviors of magnons in these compounds indicate the likelihood of cooperative excitations involving spin, lattice, as well as orbital degrees of freedom.**

PACS #: 75.30.Ds, 75.47.Lx, 75.47.Gk, 61.12.-q

**Contents**

**1. Introduction**

Half-metallic ferromagnets are characterized by completely spin-polarized electronic density of states at the Fermi level, i.e., the majority spin channel is metallic while the Fermi energy falls in a band gap in the minority spin density of states [1]. In a class of doped manganites [2] which exhibit the colossal magnetoresistance (CMR) effect [3] — the extremely large drop in resistivity induced by application of a magnetic field near the Curie temperature ($T_C$), the FM metallic state has been suggested theoretically [4] and experimentally [5] as a possible half-metallic state.

The revival in the study of manganites has led to the observation of a large array of emergent phase structures and transitions [6, 7]. It is believed that the richness of physical properties is resulted from the multitude of competing ground states — the equilibrium between phases is very subtle and small perturbations may induce a large response, which can be tuned by chemical doping, structural manipulation, strain induction, or the application of external stimuli, such as pressure, electric and magnetic fields, etc. In general, the fundamental physics behind these emergent phenomena is related to the complexity which is associated with strong interplay between charge, spin, orbital, and lattice.

The metallic ground state associated with FM order in doped manganites has originally understood by the DE interaction model [8]. In this model, the kinetics of itinerant electrons in these materials strongly correlates with localized spins in the Mn sites through the strong Hund's rule coupling. The electron hopping maintains its optimal manner when the net spins of Mn sites are all parallel. Consequently, FM ordering of the localized spins promotes metallic state with high conductivity of electrons, and vice versa. While paramagnetic (PM) order prevents electrons from hopping thus endorses insulating state. Although the DE interaction has been recognized as a basic ingredient for the coupled FM metallic to paramagnetic (PM) insulator transition as well as the CMR effect, the nature of the FM-metallic ground state is still not understood [6]. Especially, as we will focus on in this review, the spin dynamics in the FM metallic manganites is by *no means conventional*. The unconventional behaviors of spin dynamics in FM manganites are revealed by the deviation of the dispersion, the linewidth, and the long-wavelength stiffness of magnons from the expectations of the simple DE Model. Based upon the fact that a strong interplay exists between different degrees of freedom and their excitations several theoretical approaches beyond the simple DE model have been attempted. These include considering the magnon-phonon coupling, effects of electron-electron correlation, orbital fluctuations, and local phase inhomogeneities. Yet it is fair to conclude that none of the prevailing models can account for the observed magnon behaviors.

The paper is organized as follows. In section 2 we describe the magnons in a canonical DE FM system with a strong Hund's rule coupling including expected magnon dispersion, lifetime, and stiffness. Section 3 contains a brief description of inelastic neutron scattering (INS) as an ideal probe to measure the magnon properties. The results of magnon measurements from relatively high-$T_C$ or large bandwidth manganites are reviewed in Sec. 4. Section 5 and 6 present the results of magnon measurements from low-$T_C$ compounds with strong Jahn-Teller (J-T) and other correlation effects where the unusual magnon behaviors were observed. In Sec. 7 we discuss magnon damping and possible correlation with lattice dynamics. Section 8 repots results on the incoherent spin dynamics when temperature approaches to $T_C$ and the possible correlations with phase separation. Some theoretical approaches in account for these observed mangon behaviors, especially the zone boundary magnon softening, are discussed in Sec. 9. A brief summery is given in Section 10.

**2. Magnons in DE ferromagnet**

In this review, we concentrate on perovskite manganites with a transition from a high temperature paramagnetic (PM) insulator to a low temperature FM metal at $T_C$, mainly pseudo-cubic perovskite manganites $R_{1-x}A_xMnO_3$ (e.g. R = La, Nd, Pr, A = Sr, Ca, Pb) (see Fig. 1). The compounds that exhibit this behavior have been partially hole-doped away from a parent antiferromagnetic (AF) insulator $RMnO_3$ by divalent substitution on the cation site, such as $La_{0.7}Ca_{0.3}MnO_3$. The Mn $3d$ levels, split by the oxygen octahedral crystal field to a lower energy $t_{2g}$ triplet and a higher energy $e_g$ doublet, are filled according to Hund's rule such that all spins are aligned on a given site by a large intra-atomic exchange $J_H$. Electronic conduction arises from the hopping of an electron from $Mn^{3+}$ to $Mn^{4+}$ with electron transfer energy $t$. In general, these systems can be treated as a single $e_g$ band of electrons interacting with localized core spins in $t_{2g}$ triplet by a Hund rule exchange interaction and described by Kondo-type lattice model [9 - 11]:

$$H = -t \sum_{<i,j>,\alpha} (c_{i\alpha}^+ c_{j\alpha} + c_{j\alpha}^+ c_{i\alpha}) - \frac{J_H}{2S} \sum_{i,\alpha,\beta} \vec{S}_i \cdot \vec{\sigma}^{\alpha\beta} c_{i\alpha}^+ c_{i\beta} \qquad (1)$$

where $c_{j\sigma}$ is the fermionic operator corresponding to conduction electrons, hopping between the atomic sites of magnetic Mn ions with spins $\vec{S}_i$ ($S = 3/2$), and the vector $\vec{\sigma}^{\alpha\beta}$ is composed of Pauli matrices. In the limit of $J_H \gg t$, the itinerant conduction electrons must be locally align with the core spins on any site such that the ground state is a FM state. The ferromagnetic interaction between core spins mediated by conduction electrons is referred as DE model [8]. In the canonical limit $t/J_H \to 0$ and large-$S$ approximation [12], the Kondo-type model is equivalent to the nearest-neighbor Heisenberg ferromagnet which is generally described by

$$H = -\sum_{ij} J_{ij} \vec{S}_i \cdot \vec{S}_j \qquad (2)$$

with coupling $J_{ij}$ between pairs of spin at site $\vec{R}_i$ and $\vec{R}_j$. For a FM ground state and in a linear approximation, the corresponding magnon dispersion is given by

$$\hbar\omega(\vec{q}) = \Delta + 2S[J(\vec{0}) - J(\vec{q})] \qquad (3)$$

where $\vec{q}$ is the momentum transfer (or momentum transfer in the first Brillouin zone or reduced vector) during the magnon excitations, $\Delta$ is the magnon energy gap representing the energy to uniformly rotate the entire spin system away from easy direction of magnetization (thus sometimes called the magnetic anisotropy gap) and

$$J(\vec{q}) = \sum_j J_{ij} \exp[i\vec{q}\cdot(\vec{R}_i - \vec{R}_j)] \qquad (4)$$

for a Bravais lattice. For a pseudo-cubic crystal structure of $R_{1-x}A_xMnO_3$, the first few neighbor exchanging couplings are schematically shown in Fig. 2. In addition to the nearest-neighbor interaction with exchange coupling constant $J_1$ along the equivalent [1,0,0], [0,1,0] or [0,0,1] direction, $J_2$ represents the next nearest-neighbor exchange coupling, thus is the coupling in the planar [1,1,0] direction. $J_3$ is the coupling in the cubic diagonal [1,1,1] direction. $J_4$ is the next neighbor coupling in along the same directions $J_1$. In the long wavelength limit $\vec{q} \to 0$, Eq. 3 reduces to

$$\hbar\omega(\vec{q}) = \Delta + Dq^2 \qquad (5)$$

in which

$$D = \frac{S}{3}\sum_j J_{ij}|\vec{R}_i - \vec{R}_j|^2 \qquad (6)$$

is defined as the spin stiffness for a pseudo-cubic system. For the simplest case where only the nearest-neighbor coupling is considered, the magnon dispersion Eq. 3 further reduces to [10]

$$\hbar\omega(\vec{q}) \cong \Delta + 4J_1 S[3 - \cos(q_x a_0) - \cos(q_y a_0) - \cos(q_z a_0)] \qquad (7)$$

where $J_1$ is the nearest-neighbor exchange constant. Here we define $(q_x, q_y, q_z) \equiv (\frac{2\pi}{a_0}h, \frac{2\pi}{a_0}k, \frac{2\pi}{a_0}l)$. For a cubic system, the spin stiffness shown in Equation (6) can be further simplified as

$$D = 8\pi^2 SJ_1 \qquad (8)$$

in the reciprocal-lattice units (*rlu*). In the mean filed Heisenberg model [13], the Curie temperature is directly proportional to the exchange constant [12],

$$T_C = \frac{4S(S+1)}{k_B} J_1. \quad (9)$$

It should be noted that, in practice, any quantum fluctuation effect tends to reduce the effective $T_C$ even in the Heisenberg ferromagnet. The magnon bandwidth is also determined by

$$W_{sw} \equiv \hbar\omega[\vec{q} = (\frac{1}{2}, \frac{1}{2}, \frac{1}{2})] = 24SJ_1. \quad (10)$$

Under this scenario, the spin-wave stiffness $D$ and bandwidth should be linearly proportional to the Curie temperature ($T_C$). In a mean-field theory for a Heisenberg FM case, $T_C$ so as $J_1$ is proportional to the average kinetic energy ($t$) [9 - 12], i.e., $D \propto J_1 \propto t \propto T_C$.

In this simple Heisenberg ferromagnet, the spin waves are the exact eigenstates of Heisenberg Hamiltonian. At $T = 0$ K, magnons are non-interacting quasiparticles with long lifetime and no damping.

### 3. Neutron as a probe for magnon excitations

Neutron scattering has been a vital tool in probing both magnetic ordering and spin dynamics like magnon excitations [14, 15]. For unpolarized inelastic scattering neutrons with momentum transfer vector $\vec{Q} = \vec{k}_i - \vec{k}_f = \vec{G} + \vec{q}$ where $\vec{G}$ is the reciprocal-lattice vector and energy transfer $\hbar\omega = \frac{\hbar^2}{2m_n}(k_i^2 - k_f^2)$, the differential cross section for the scattering from a system of electron spins is given by

$$\frac{d^2\sigma}{d\Omega_f dE_f} = \frac{N}{\hbar} \frac{k_f}{k_i} p^2 e^{-2W} \sum_{\alpha\beta} (\delta_{\alpha\beta} - \hat{Q}_\alpha \hat{Q}_\beta) S^{\alpha\beta}(\vec{Q}, \omega) \quad (11)$$

with the scattering function

$$S^{\alpha\beta}(\vec{Q}, \omega) = \frac{1}{2\pi} \int_{-\infty}^{\infty} dt e^{-i\omega t} \sum_l \langle S_0^\alpha(0) S_l^\beta(t) \rangle \quad (12)$$

where $\langle ... \rangle$ denotes an average over configurations. For magnetic ions, the amplitude for magnetic scattering is given by $p = (\frac{\gamma r_0}{2}) g f(\vec{Q})$ where $\gamma = 1.913$ is the neutron gyromagnetic ratio, $r_0$ the classical electron radius, and $f(\vec{Q}) = \int \rho_s(\vec{r}) e^{i\vec{Q}\cdot\vec{r}} dr$ the magnetic form factor which is the Fourier transform of the normalized unpolarized spin density $\rho_s(\vec{r})$ on an atom. For magnon scattering, a neutron, scattering from a magnetic system, can adsorb or emit one or more magnons. For a single magnon process with collinear spins aligned parallel to z-axis such that magnons involve $S^x$ and $S^y$, then summation in Eg. 11 can be expressed as

$$\sum_{\alpha\beta}(\delta_{\alpha\beta} - \hat{Q}_\alpha \hat{Q}_\beta) S^{\alpha\beta}(\vec{Q},\omega) = \frac{1}{2}(1+Q_z^2) S_{sw}(\vec{Q},\omega) \quad (13)$$

where $S_{sw}(\vec{Q},\omega)$ is the inelastic scattering function for magnons. In particular, for simple Heisenberg ferromagnetic with small $\vec{q} = \vec{Q} - \vec{G}$ where magnons exhibit parabolic-type dispersion, the inelastic scattering function can be further simplified as

$$S_{sw}(\vec{Q},\omega) = S \sum_{\vec{G},\vec{q}} [(n_{\vec{q}}+1)\delta(\vec{Q}-\vec{q}-\vec{G})\delta(\omega-\omega_{\vec{q}}) + n_{\vec{q}}\delta(\vec{Q}+\vec{q}-\vec{G})\delta(\omega+\omega_{\vec{q}})]. \quad (14)$$

The essential information contained in neutron scattering is that neutron scattered with momentum and energy transfer $\vec{Q}$ and $\hbar\omega$ directly probe a single Fourier component of the spin-pair correlation function (i.e., the scattering function). For a given energy, the scattering function as a function of $\vec{Q}$ provides the information of the dynamic spin-spin correlations. In reality, magnon-magnon interaction and other channels of interactions (like magnon-phonon) lead to magnon damping [15]. To include the effect of damping it is convenient to make use of the fluctuation-dissipation theorem to relate the scattering function to the imaginary part of the generalized susceptibility [15, 16]:

$$S^{\alpha\beta}(\vec{Q},\omega) = [n(\omega)+1] \operatorname{Im} \chi^{\alpha\beta}(\vec{Q},\omega) \quad (15)$$

where $n(\omega)+1 = [1-\exp(-\frac{\hbar\omega}{k_B T})]^{-1}$ is the Bose population factor. Therefore, the neutron scattering directly measures $\operatorname{Im} \chi^{\alpha\beta}(\vec{Q},\omega)$, hence providing the information of magnon damping due to the existence of different interactions. For example, in the damped simple harmonic oscillator (DSHO) approximation [16], the normalized dynamic susceptibility Im $\chi^{\alpha\beta}(\vec{Q},\omega)$ can be expressed as

$$\text{Im}\,\chi(\vec{Q},\omega) = \frac{4\gamma\omega\omega_0}{\pi[(\omega^2-\omega_0^2)^2 + 4(\gamma\omega)^2]} \qquad (16)$$

where $\gamma$ characterizes the magnon damping while $\omega_0$ is associated with the magnon dispersion relation.

The experimental results presented below were mostly obtained from the inelastic neutron scattering of single crystal manganites. Most of experiments were performed on triple-axis neutron scattering spectrometers except otherwise indicated. The manganite crystals for the experiments were mainly grown by the traveling solvent floating zone technique. The reciprocal-lattice units (*rlu*) is used to label wave vectors so that the momentum transfer $(q_x,q_y,q_z)$ in units of Å$^{-1}$ are at reciprocal space positions $(H,K,L) = (q_x a_x/2\pi, q_y a_y/2\pi, q_z a_z/2\pi)$ *rlu*, where $a_x$, $a_y$, $a_z$ are the lattice parameters. For simplicity, we label all wave vectors in terms of the pseudo-cubic unit cells with lattice parameter *a*. In reality, most of manganites have a lower-symmetry structure such as orthorhombic one which is slightly distorted from the cubic lattice (see the ball structure model for an orthorhombic phase in Fig. 1). In the pseudocubic perovskite unit cell, $a_x = a_y = a_z = a$. In this notation, the zone boundary along the [$\xi,0,0$], [$\xi,\xi,0$], and [$\xi,\xi,\xi$] directions for FM magnons are at the (0.5,0,0) *rlu*, (0.5,0.5,0) *rlu*, and (0.5,0.5,0.5) *rlu*, respectively.

## 4. Magnons in high-$T_C$ manganites

The earlier experiments on the measurement of magnon excitations were carried out from the crystals of La$_{0.7}$Pb$_{0.3}$MnO$_3$ by Perring *et al.* [17], La$_{0.7}$Sr$_{0.3}$MnO$_3$ by Martin *et al.* [18] and La$_{0.8}$Sr$_{0.2}$MnO$_3$ by Endoh *et al.* [19]. For these compounds with relatively higher Curie temperature ($T_C$ = 355 K for La$_{0.7}$Pb$_{0.3}$MnO$_3$, 378 K for La$_{0.7}$Sr$_{0.3}$MnO$_3$, and 312 K for La$_{0.8}$Sr$_{0.2}$MnO$_3$), the dispersion of magnon along all three high-symmetry directions, [1,0,0], [1,1,0], and [1,1,1], was explained by simple Heisenberg model with solely a nearest neighbor coupling. As shown in Fig. 3, the dispersion relation using Eq. 7 is entirely sufficient to account for the data obtained from La$_{0.7}$Pb$_{0.3}$MnO$_3$ at 10 K [17], giving the spin-wave stiffness $D \cong 133.7$ meVÅ$^2$ with $2J_1S = 8.79 \pm 0.21$ meV and $\Delta = 2.51$

± 0.46 meV. The value of the gap $\Delta$ was obtained by fitting the dispersion data to the model instead of that from direct measurement. It was found that adding the second and third nearest-neighbor exchange interactions does not improve the fit. The simple nearest-neighbor FM Heisenberg model, with a consideration of the effect of fluctuations [13], also accounts for the estimate of $T_C$ of the materials to within 15%. On the other hand, the data for the low-energy and near zone boundary magnon excitations are still lacking. A more complete data set to map the magnon dispersion and extract the value of spin-wave stiffness $D$ is desirable.

The results of the magnon dispersion in long-wavelength for $La_{0.7}Sr_{0.3}MnO_3$ along the [1,1,0] direction measured by Martin *et al.* [18] also suggested the dispersion can be understood by the simple Heisenberg model. The fit to the data obtained at 10 K for the dispersion gave the spin-wave stiffness $D = 188 \pm 8$ meV Å$^2$ and a very small fitting gap $\Delta = 0.75 \pm 0.40$ meV. Vasiliu-Doloc *et al.* [19] obtained $D = 176 \pm 5$ meV Å$^2$ for $La_{0.7}Sr_{0.3}MnO_3$ and $166.8 \pm 5$ meV Å$^2$ for $La_{0.8}Sr_{0.2}MnO_3$ at 15 K although the earlier and less accurate measurements [20] gave a lower value of $D$. However, no energy gap (at least $\Delta < 0.02$ meV which is within the instrumental energy resolution) was measured. From this value of $D$ one would be able to calculate the mean-field value of $T_C$ based upon Eq. 8 and Eq. 9. It was found [18] that the calculated $T_C$ value is more than twice higher than the actual $T_C = 378$ K of the system. This has been used as an indication for the itinerant character of the system, since an itinerant ferromagnet generally has a lower $T_C$ compared to the mean-field $T_C$ value but large $D$ value [20]. Obviously this argument assumes that magnons follow completely the cosine-like dispersion (Eq. 7) such that a large $D$ value would have a large magnon bandwidth. Unfortunately, the magnon dispersion was mapped only in low-q range while the exact magnon bandwidth was not clear for this system. A complete magnon dispersion to the zone boundary along the [1,0,0] direction has recently reported [21] but a zone boundary softening deviated from the nearest-neighbor Heisenberg model has been observed. We will discuss in the next section.

The experimental results on the magnon dispersion and the doping (x) dependence of spin-wave stiffness for $La_{1-x}Sr_xMnO_3$ ($x \leq 0.3$) seems further confirming the validity of nearest-neighbor Heisenberg model. Based upon the long-wavelength part of the magnon dispersion, Endoh and Hirota [20] discovered that the *x*-dependence of *D* almost completely coincided with that of $T_C$, both solely depending on $J_1$, thus following the simple relationship $D \propto J_1 \propto T_C$ as predicted by nearest-neighbor Heisenberg model. Figure 4 summarized the correlation between measured transition temperature (either the Curie or Neel temperature [22]) and spin-wave stiffness as a function of doping concentration of $La_{1-x}Sr_xMnO_3$ from these reported by Ref. [20] and other measurements [18, 19, 23]. Moreover, the ratio between the calculated $T_C$ based upon Eq. 9 by using experimentally determined $J_1$ and the actual $T_C$ is almost the same in the doping range studied ($x \leq 0.3$). However, it is known that there is a FM metal to FM insulator transition below $x$ = 0.175. Thus the independence of the ratio on doping has been speculated that the electron correlation energy remains essentially unchanged across the metal-to-insulator transition of $La_{1-x}Sr_xMnO_3$.

### 5. Zone boundary magnon softening

Evidence of the magnon behavior deviating from the simple nearest-neighbor Heisenberg model was first discovered in $Pr_{0.63}Sr_{0.37}MnO_3$ with $T_C$ = 301 K [24]. A clear softening of the magnon dispersion at the zone boundary for $T < T_C$ and significant broadening of the zone boundary magnons as $T \to T_C$ have been observed. Figure 5 shows the magnon dispersion of $Pr_{0.63}Sr_{0.37}MnO_3$ along the three high-symmetry directions. It can be seen that only the nearest-neighbor coupling is *not enough* to account for the dispersion of magnon. The solid line in Fig. 5 is the result of a fit to only nearest-neighbor interactions for the small-*q* range ($\xi$ < 0.2), resulting in $\Delta$ = 1.3 ± 0.3 meV and $2J_1S$ = 8.2 ± 0.5 meV. Though fitted results for the small-*q* range are similar to those obtained from the dispersion in $La_{0.7}Pb_{0.3}MnO_3$, a large deviation by 15-20 meV near the zone boundary ($\xi$ = 0.5 in cubic structure) is evident, in sharp contrast with that in $La_{0.7}Pb_{0.3}MnO_3$. The spin wave stiffness *D* of $Pr_{0.63}Sr_{0.37}MnO_3$ is 165 meV Å$^2$ [24].

A Heisenberg model including higher-order couplings to fourth neighbor interactions has been taken to fit the full data set and the results are shown in Fig. 5 (dashed curves) [24]. This fit gives $\Delta$ = 0.2 ± 0.3 meV, $2J_1S$ = 5.58 ± 0.07 meV, $2J_2S$ = -0.36 ± 0.04 meV, $2J_3S$ = 0.36 ± 0.04 meV, and $2J_4S$ = 1.48 ± 0.10 meV. It was also found that a better fit for data near the zone boundary in the [0,0,1] direction required the next Fourier term ($J_8$) as compared with that in the [1,1,0] and [1,1,1] direction. Though $J_2$ and $J_3$ were necessary to fit the data, the fourth nearest-neighbor coupling $J_4$ is particularly important to correct the nearest-neighbor coupling. It seems that the long range and nonmonotonic behavior of $J(\bar{q})$ required by the measured data rules out a simple Heisenberg Hamiltonian with nearest-neighbor exchange coupling.

Such a strong zone boundary magnon softening has been further confirmed by other measurements on several manganites with relatively low $T_C$ [25 - 28]. Compared with $Pr_{0.63}Sr_{0.37}MnO_3$ ($T_C$ = 301 K), $La_{0.7}Ca_{0.3}MnO_3$ ($T_C$ = 238 K) and $Nd_{0.7}Sr_{0.3}MnO_3$ ($T_C$ = 198 K) have much lower $T_C$ though all three compounds have a FM metallic ground state, have an identical $T$-dependence of resistivity, and exhibit a metal-to-insulator transition around $T_C$ [see Fig. 6]. In particular, $La_{0.7}Ca_{0.3}MnO_3$ is a widely studied manganite with the optimized doping level for the CMR effect [29]. One advantage of using $La_{0.7}Ca_{0.3}MnO_3$ to study the magnon behavior is that La is not a magnetic ion so that no excitation due to the crystal electric field (CEF) level involves in magnon excitations. Dai *et al*. [27] have carried out the detailed studies on the magnon dispersion, damping, as well as its temperature dependence. Figure 7 presents a complete set of constant-*q* scans in the [1,0,0] direction (the same as the [0,0,1] direction in the notation for cubic perovskite structure through this paper) for the magnon excitations of $La_{0.7}Ca_{0.3}MnO_3$ at 10 K. The magnon peaks are well resolved up to the zone boundary without any other magnetic excitations (like those due to CEF) in the observed energy window. However, a large increase in linewidth and decrease in intensity near the zone boundary are evident in the magnon excitation spectra. Figure 8 presents the magnon dispersions of $La_{0.7}Ca_{0.3}MnO_3$ accompanied with $Nd_{0.7}Sr_{0.3}MnO_3$ and $Pr_{0.63}Sr_{0.37}MnO_3$ along both [0,0,1] and [1,1,0] directions. Remarkably, the magnon dispersions of these three manganites are almost identical at the measured energies, showing a large zone boundary softening in both [0,0,1] and [1,1,0] directions. This indicates that the magnetic exchange coupling strength is insensitive to large difference (more than 100 K) of $T_C$'s, in sharp contrast to the prediction of $D \propto J_1 \propto t \propto T_C$. *These are the conclusive evidence that the spin dynamics in these manganites can not be explained by simple nearest-neighbor Heisenberg model.*

Endoh *et al.* [28] have measured the magnon excitation of the FM state of $Sm_{0.55}Sr_{0.45}MnO_3$ which has much lower $T_C$ (~ 135 K) and is located on the verge of a doping-induced metal-insulator transition [30]. The anomalous zone boundary magnon softening has also been observed in this low-$T_C$ material along the [1,0,0], [1,1,0] and [1,1,1] directions. In particular, an anisotropic softening was observed with the largest softening in the [1,0,0] direction. Yet, no obvious broadening of the magnon spectra has been observed near zone boundary where the dispersion tends to show softening, in sharp contrast to these observed in $Pr_{0.63}Sr_{0.37}MnO_3$, $La_{0.7}Ca_{0.3}MnO_3$, and $Nd_{0.7}Sr_{0.3}MnO_3$. The magnon dispersions of $Sm_{0.55}Sr_{0.45}MnO_3$ have been fitted to the Heisenberg model with the nearest-neighbor ($J_1$) and fourth-neighbor ($J_4$) couplings. However, the fit for the zone boundary dispersion along the [1,1,0] and [1,1,1] directions are not as good as that along the [1,0,0] direction [28] thus giving the possible uncertainty in the determination of $J_4$. Nevertheless, it is found that the ratio $J_4/J_1$ (~ 0.6) is much larger than these obtained from other compounds (see Table 1).

In contrast with the results for high-$T_C$ manganites as we discussed in the previous section, the zone-boundary softening has been reported in $La_{0.68}Ba_{0.32}MnO_3$ [31] and $La_{0.7}Ba_{0.3}MnO_3$ [32] which also have relatively high-$T_C$. A fit of the reported data [31] for $La_{0.68}Ba_{0.32}MnO_3$ ($T_C$ = 336 K) indicates a non-zero fourth-neighbor coupling with $2SJ_4$ = 1.59 meV (see table 1). Chatterji *et al.* [32] have measured the magnon dispersions of $La_{0.7}Ba_{0.3}MnO_3$ ($T_C$ = 350 K) at 1.5 K along the [1,0,0] and [1,1,0] directions and determined the spin-wave stiffness $D$ = 152 ± 3 meV Å$^2$ by fitting the data to the Heisenberg model. However, a large deviation of the fitting curves from the experimental dispersions was found near the zone boundary. The magnons show zone boundary softening and are heavily damped for higher $q$ with larger linewidths than the instrumental resolution. In order to fit the dispersion data in the whole $q$ range, higher order terms in the Heisenberg model is needed to take into account (see Table 1), in contrast with these obtained from $La_{0.7}Pb_{0.3}MnO_3$ [17]. So far it is still an open issue that whether or not the zone-boundary magnon softening and damping is the generic features of all FM manganites. As we have mentioned above, more measurements are needed on $La_{0.7}Pb_{0.3}MnO_3$ to determine the entire dispersion curve. There may not be enough data at present to conclude that the magnon behavior in $La_{0.7}Pb_{0.3}MnO_3$ or other high-$T_C$ manganites can indeed be described by a simple Heisenberg model with nearest neighbor exchange interaction.

### 6. Doping-dependence of magnon excitations

Based upon the results about the magnon dispersion described above, many important issues need to be addressed: *How the observed softening correlate with the carrier concentration (x), on-site disorder, and strength of lattice distortion*? So far, it seems quite clear that the zone boundary softening occurs in these relatively low-$T_C$ or narrow-band materials though it is not quite clear for the high-$T_C$ compounds. If indeed that the softening as well as the unusual magnon damping mainly occurs in these low-$T_C$ manganites which has large John-Teller effects, it may be associated with strong spin-lattice/orbital couplings.

In order to further gain insight into the issue of zone boundary softening, Ye *et al*. [21] recently have systematically analyzed existing magnon data and taken additional data in the FM metallic state of $R_{1-x}A_xMnO_3$ at judicially selected doping levels. In additional to the single crystals of $La_{0.75}Ca_{0.25}MnO_3$, $La_{0.7}Ca_{0.3}MnO_3$, $Nd_{0.7}Sr_{0.3}MnO_3$, $Pr_{0.63}Sr_{0.37}MnO_3$, $La_{0.7}Sr_{0.3}MnO_3$, $Sm_{0.55}Sr_{0.45}MnO_3$, $La_{0.7}Ba_{0.3}MnO_3$, and $La_{0.68}Ba_{0.32}MnO_3$, which have FM metallic phase as the ground state, $Pr_{0.7}Ca_{0.3}MnO_3$ and $Pr_{0.55}(Ca_{0.85}Sr_{0.15})_{0.45}$ $MnO_3$ have also been used for the study. The later two samples exhibits AF/canted AF insulating ground state but can be tuned into FM metallic state by applying an external magnetic field (see the phase diagram [21, 33] of these two compounds in Fig. 9), thus the magnon behavior in the field-induced FM metallic state can also be studied. Though application of magnetic field adds a field-induced Zeeman gap [34], it is remarkable that all three samples exhibit very similar low-$q$ behavior disregarding the difference in achieving the FM metallic states either by temperature or by magnetic field. As shown in the inserts of Fig. 10 [21], the slopes of the magnon energy (*E*) versus $q^2$ lines yield *D* values of 150 ± 3, 145 ± 8, and 152 ± 3 meV Å$^2$ for $La_{0.75}Ca_{0.25}MnO_3$, $Pr_{0.7}Ca_{0.3}MnO_3$ and $Pr_{0.55}(Ca_{0.85}Sr_{0.15})_{0.45}$ $MnO_3$, respectively. This clearly indicates that the spin-wave stiffness is independent of how the FM metallic phase is realized or the carrier concentration.

To determine the evolution of magnon excitations in $R_{1-x}A_xMnO_3$ as a function of doping, Fig. 11 (a) summarizes the magnon dispersions along the [1,0,0] direction for a series of $R_{1-x}A_xMnO_3$ with $x \approx 0.3$ [19, 21, 25, 27, 31] while Fig. 11(b) presents the dispersions along the same direction but for different doping concentrations [21, 24, 28]. The solid curves in the figure are phenomenological fits to the data using the Heisenberg Hamiltonian Eq. (3) and (4) with nearest-neighbor ($J_1$) and fourth-nearest-neighbor ($J_4$) exchange coupling. In the low-$q$ limit, $E(q) = \Delta + 8\pi^2 S(J_1 + 4J_4)q^2$, instead of using Eq (5) which takes only the nearest-neighbor coupling. It is found [21, 24, 28] that the contributions from the second-nearest-neighbor ($J_2$) and third-nearest-neighbor ($J_2$) exchange coupling are negligible. While the magnons show similar dispersion for $R_{1-x}A_xMnO_3$ with $x = 0.3$ (Fig. 11a), the doping dependence of the zone boundary magnon softening (Fig. 11b) indicates that *the higher the doping-level, the larger the zone boundary softening*.

We should emphasize that besides the doping, the effect of *A*-site disorder (or chemical disorder) arising from the mismatch between rare- and alkaline-earth-metal ions might induce anomalous spin dynamical behavior [35 - 38]. The *A*-site disorder is characterized by the standard deviation of the ionic radii: $\sigma^2 = \sum_i (x_i r_i^2 - \bar{r}^2)$ where $x_i$ is the fractional occupancies of *A*-site species, $r_i$ and $\bar{r} = \sum_i x_i r_i$ are the individual and averaged ionic radius, respectively [39]. Figure 12 summarizes the $\sigma^2$- and $\bar{r}$-dependence of the spin-wave stiffness $D$, $J_1$, and the ratio of $J_4/J_1$ which manages the zone boundary softening [21]. Surprisingly, varying disorder seems to have *no* systematic effect on $D$ and $J_1$. With increasing disorder, the spin-wave stiffness falls within a bandwidth of $D = 160 \pm 15$ meV Å$^2$ and the nearest-neighbor exchange coupling falls within a bandwidth of $2SJ_1 = 7$ meV (Fig. 12 a and b). Furthermore, the ratio of $J_4/J_1$ show *no* dependence on the on-site disorder (Fig 12 c), in contrast with the recent theoretical prediction [35] suggesting a significant zone-boundary softening with increasing disorder.

On the other hand, both $D$ and $2SJ_1$ show a different dependence on the average ionic radius $\bar{r}$ at *A*-sites. As shown in Fig. 12d and e, both D and $2SJ_1$ do show a parabolic curve but with a small bandwidth. This is certainly a puzzle for the understanding of the spin dynamics in FM manganites. Changing the ionic size at *A*-site will modify the length and angle of Mn-O-Mn bonds, thus leading to changes in effective transfer integral between Mn ions or the bandwidth of the electrons [40]. Despite the large change in $T_C$ by varying the average ionic radius, the kinetic energy ($D$) or the bandwidth of the electrons seems to change slightly based on these results, in consistence with earlier studies [25, 27, 41]. Moreover, $J_4/J_1$ show no dependence on $\bar{r}$ (Fig. 12 f) thus indicating that the zone boundary magnon softening is independent of $T_C$ as a general feature of the FM $R_{1-x}A_xMnO_3$ manganites.

To gain more insight into the doping dependence of spin dynamics, Fig. 13 plots the measured values of $D$, $2SJ_1$ and $J_4/J_1$ as a function of doping [21], respectively. For $La_{0.8}Sr_{0.2}MnO_2$, the value of $D = 166.8 \pm 1.51$ meVÅ$^2$ is used here from a more accurate measurement [19]. In contrast with earlier results [20] (see Fig. 4), the spin-wave stiffness $D$ keeps a value around $160 \pm 15$ meVÅ$^2$ and essentially unchanged for the doping range of $0.2 \leq x \leq 0.45$ (see Fig. 13 a) while $T_C$ varies a wide range from 135 K for $Sm_{0.55}Sr_{0.45}MnO_3$ [28] to 378 K for $La_{0.7}Sr_{0.3}MnO_3$ [19]. This is also in distinction from the theoretical predication based upon the $1/S$ spin wave expansion for DE ferromagnets by Golosov *et al*. [12].

Conversely, the nearest-neighbor exchange coupling $J_1$ and the ratio $J_4/J_1$ do show a linear-type relation with the doping concentration. $2SJ_1$ decreases while $J_4/J_1$ increases, approximately linear, with increasing $x$ (Fig. 13 b and c). This clearly shows that *the zone boundary magnon softening (denoted by the ratio $J_4/J_1$) enhances linearly with increasing doping*. However, as shown in Fig. 13 c, the $x$-dependence of ratio $J_4/J_1$ cannot be accounted by the recent proposal [28] based upon the mechanism of the $d_{3z^2-r^2}$- or $d_{x^2-y^2}$-type orbital fluctuations or free hybridized band model suggested by Solovyev *et al*. [42] (see Sec. 9 for more discussion). The change of the ratio vs. doping from the free hybridized band model is too small to account for the experiment results. While the orbital fluctuation model gives a non monotonic doping dependence of the ratio. The simple linear relation of the ratio $J_4/J_1$ to doping deserves further careful investigation.

## 7. Anomalous magnon damping

Right after the measurements on the magnon behavior in FM manganites with inelastic neutron scattering, it has been discovered [24, 26, 27] that the magnon excitation spectra have unusual large linewidths, especially near the zone boundary. As shown in Fig. 7, the magnon excitation peaks show a large increase of linewidth and damping when the reduced wave vector $\xi > 0.3$ up to the zone boundary at $\xi = 0.5$. Similar behavior has also been observed in $Nd_{0.6}Sr_{0.4}MnO_3$ [43]. Fig. 14 plots the intrinsic linewidths (FWHM) of the magnon peaks along the [0,0,1] direction for three manganites at 10 K: $Pr_{0.63}Sr_{0.37}MnO_3$ ($T_C$ = 301 K), $La_{0.7}Ca_{0.3}MnO_3$ ($T_C$ = 238 K) and $Nd_{0.7}Sr_{0.3}MnO_3$ ($T_C$ = 198 K) [27]. Near the zone center, the linewidth reaches almost the instrumental limit while a drastic increase at a wave vector larger than $\xi \sim 0.3$. More interestingly, the linewidth from all three samples shows similar behavior indicating a possible common mechanism for the effect on the magnon lifetime near the zone boundary regardless the large the difference in the Curie temperature. Meanwhile, it has been found that the strong magnon damping near zone boundary has a dependence on $T_C$, despite a systematic study is yet needed. For example, although still relatively well defined throughout the Brillouin zone in the [1,0,0] direction for both compounds, the magnon excitations are much more severely damped along the [1,1,0] direction for $Nd_{0.7}Sr_{0.3}MnO_3$ than these for $La_{0.7}Ca_{0.3}MnO_3$ [27]. Some preliminary measurements on several manganites [44] suggest that lower-$T_C$ manganites have larger zone boundary magnon damping.

Generally, the predominant effect on magnon linewidth is magnon-magnon scattering such that in the long-wavelength regime the linewidth obeys certain scaling law as $\Gamma_{mag} \propto q^4 \ln^2(k_B T / \hbar \omega_q)$ for $\hbar \omega_q \Box k_B T$, and $\Gamma \propto q^3$ for $\hbar \omega_q \Box k_B T$ [45, 23]. However, the observed sudden increase in magnon linewidth and heavy damping near the zone boundary is certainly deviated from the simple scaling law [23, 27]. Such magnon behavior should be attributed to a certain mechanism other than Heisenberg-type interactions.

At low temperature FM metallic phase, a possible mechanism account for the magnon broadening is due to the Stone continuum where magnons decays and causes electron-hole excitations. However, the FM ground state of manganite, especially if the system is indeed in the half-metallic phase, would have a complete separation of the majority and minority band due to the large Hunds-rule coupling $J_H$. As a consequence, the Stoner continuum is expected to lie at an energy scale ($2J_H$) much higher than that of the magnon excitations. It seems unlikely that the magnon broadening and damping are caused by the Stoner continuum excitations [10].

As reported by Dai *et al*. [27], the unusual magnon broadening/damping as well as softening may indicate a possible magnon-phonon coupling [25, 46, 47]. As shown in Fig. 15 (a) about the measured results from $La_{0.7}Ca_{0.3}MnO_3$, the two particular optical phonon modes ($\Omega_1$ and $\Omega_2$, respectively) which are characterized as two vibration modes associated with $MnO_6$ octahedron [48, 49] merge with the magnons in both [1,0,0] and [1,1,0] directions (see also Fig. 8). Evidently, the in-plane magnons exhibit softening and broadening when they merge with the phonons around the momentum transfer $\xi = 0.3$ for both directions. To clearly show the correlation between magnons and phonons in excitation spectrum, we plot in Fig. 14 (b) the magnon linewidth as a function of magnon energy. In the [1,0,0] direction, the magnon softens and simultaneously increases its linewidth abruptly from ~ 4 meV to ~ 12 meV at $\xi = 0.3$ where the magnon merges with the $\Omega_1$ phonon at the energy around 20 meV. In the [1,1,0] direction, the linewidth of the magnon exhibits a peak/shoulder around 20 meV where the magnon disperses across with the $\Omega_1$ phonon. Furthermore, when merging with the $\Omega_2$ phonon around $\xi = 0.3$ and the energy of 45-50 meV, the magnon damps drastically and increases its linewidth abruptly. The large error bar in the linewidth near the zone boundary in the [1,1,0] direction is *indeed* because of the low peak intensity in the excitation spectra due to the significant magnon damping when the magnon probably entangles with the $\Omega_2$ phonon branch. Furthermore, such drastic magnon damping close to the zone boundary is much more enhanced for lower $T_C$ samples and in the [1,1,0] than [1,0,0] direction, indicating *an anisotropy of magnon damping* even in the $MnO_2$ plane [23, 27, 44]. Actually, in the case of $Nd_{0.7}Sr_{0.3}MnO_3$ which has lower $T_C$ than $La_{0.7}Ca_{0.3}MnO_3$, the zone boundary magnons are overdamped to be experimentally measured [27].

It is worthy to mention that the $\Omega_2$ phonon merging with the [1,1,0] magnon branch is a *JT-active* mode associated with the oxygen vibration in the $MnO_6$ octahedron while the $\Omega_1$ phonon is an *external* mode associated with the La vibration against the $MnO_6$ octahedron. If spin-phonon interaction is responsible for the broadening/damping of magnon, the difference in the character of these two phonon modes should directly relate to the observed difference in damping of the in-plane magnons between the [1,0,0] and [1,1,0] direction. In contrast, in the [1,1,1] direction, the magnon shows no obvious and unusual behavior [see the inset of Fig. 15 (b)]. These results indicate that the anomalous broadening and damping behavior of the in-plane but not the out-of-the-plane magnons occurs when the magnons and optical phonons merge in the energy-momentum space.

Nevertheless, more quantitative measurements on the intrinsic linewidth of magnon excitations are clearly needed. Even for the magnon behavior in the [1,0,0] direction which has been most widely studied, the measured results on linewidths near the zone boundary is still controversial. The results measured from $Sm_{0.55}Sr_{0.45}MnO_3$ [28] show the magnon linewidth along the [1,0,0] direction is within 0.7 meV (instrumental energy resolution limited), more than one order smaller than the results from other groups [23, 24, 27]. It was claimed that there is neither anomalous broadening of the magnon spectra nor loss of the scattering intensities in $q_s$ where the dispersion tends to show softening, thus excludes the level crossing with phonons or the phases separation. In order to identify the nature for the finite lifetime of magnons, especially that close to the zone boundary, polarized neutron scattering experiment is ideal which allow completely separating the contribution from other excitations like phonons [50].

## 8. Temperature Dependence and Incoherent spin dynamics near $T_C$

The $T$-dependence of spin dynamics in the FM metallic manganites is reflected by the evolution of the magnon dispersion including the spin-wave stiffness as a function of temperature as well as some unusual incoherent spin dynamics as $T \to T_C$. The earlier measurement on $Pr_{0.63}Sr_{0.37}MnO_3$ (see Fig. 5) [24] shows that the magnon dispersion relation uniformly softens with increasing temperature. Fig. 16 shows the measured $T$-dependence of magnon dispersion along the [1,0,0] direction from $La_{0.7}Ca_{0.3}MnO_3$, further confirming the gradually softening with increasing temperature. However, the temperature has non-uniform effects on the magnon lifetime [23, 24]. While there is no obvious effect near the zone center, magnons near zone boundary show substantially increase in linewidth and decrease in intensity with increasing $T$. For example, as shown in Fig. 17, the zone boundary linewidths in $Pr_{0.63}Sr_{0.37}MnO_3$ along the [1,0,0] direction are nearly doubled from their value of $8.4 \pm 0.5$ meV at $T = 10$ K to $13.2 \pm 1.9$ meV at $T = 265$ K. For La$_{0.7}$Ca$_{0.3}$MnO$_3$ which has a lower $T_C$ compared with Pr$_{0.63}$Sr$_{0.37}$MnO$_3$, the magnons near zone boundary are overdamped when $T \to T_C$ such that no reliable dispersion data can be obtained along the [1,0,0] direction when $\xi \geq 0.3$. Such $T$-induced broadening and damping near zone boundary are even severe in the [1,1,0] direction and for lower $T_C$ manganites [44]. This indicates that there is other effect causing the enhanced magnon damping near zone boundary rather than the simple magnon-magnon scattering and such effects should be $T_C$-dependent.

Even in the long-wavelength (low-$q$) limit, the $T$-dependence of magnons also shows unusual behavior which is clearly dependent on $T_C$ [19, 23, 25, 26, 51, 52]. It is found [25, 26] that, for low-$T_C$ samples, the spin-wave stiffness $D(T)$ exhibits a power law behavior as a function of temperature but does *not* collapse as $T \to T_C$, thus challenge the simple theories based on a Heisenberg ferromagnetism and DE model. For a Heisenberg ferromagnet, $D(T)$ is expected to follow mode-mode coupling theory [53] with $D(T) = D(0)(1 - AT^{5/2})$ at low $T/T_C$. As $T \to T_C$, $D(T)$ should renormalize to zero at $T_C$ as power law behavior like $[(T - T_C)/T_C]^{\nu - \beta}$ with $\nu - \beta = 0.34$ [54]. Figure 18 presents the measured $D(T)$ vs $T/T_C$ for three samples: Nd$_{0.7}$Sr$_{0.3}$MnO$_3$ ($T_C$ = 198 K), La$_{0.7}$Ca$_{0.3}$MnO$_3$ ($T_C$ = 238 K) and Pr$_{0.63}$Sr$_{0.37}$MnO$_3$ ($T_C$ = 301 K). The measured $D(T)$ for Pr$_{0.63}$Sr$_{0.37}$MnO$_3$ almost follows the theoretical expectation from a Heisenberg ferromagnet [25]. However, for La$_{0.7}$Ca$_{0.3}$MnO$_3$ and Nd$_{0.7}$Sr$_{0.3}$MnO$_3$, which have lower $T_C$, it seems that $D(T)$ shows no evidence of the magnon collapse at $T_C$ although the magnetization $M(T)$ of these two compounds does not show unusual behavior.

To further characterize the spin dynamics of FM manganites when $T \to T_C$ several groups have studied the spin diffuse scattering near $T_C$ [19, 23, 25, 26, 51, 52, 55, 56]. An anomalous and field-dependent central diffusive component which develops above $T \sim 0.8$ $T_C$ for La$_{0.7}$Ca$_{0.3}$MnO$_3$ [25] and $T \sim 0.9$ $T_C$ for Nd$_{0.7}$Sr$_{0.3}$MnO$_3$ [23] and dominates the fluctuation spectrum as $T \to T_C$ [25, 23] has been observed for the low-$T_C$ samples, *coincided* with the non-collapse behavior of $D(T)$. This central component is the result of quasi-elastic spin diffuse scattering. Figure 19 presents the $T$- and field-dependence of the central diffusive component as well as the magnon peaks measured from La$_{0.67}$Ca$_{0.33}$MnO$_3$. The central component decreases while the magnon component increases in intensity with increasing field, thus the strength of the spectrum shifts from the central component into the magnon one as the field is increased. Meanwhile, it is found that [25, 26, 51, 52, 56] that the temperature at which the central component appears is related to $T_C$. In $Pr_{0.63}Sr_{0.37}MnO_3$ the central component emerges only when $T > 0.95\ T_C$ [25], thus much close to $T_C$.

This anomalous central component has been interpreted [26] as the signature of the formation of spin polarons similar to the ferromagnetic "droplets" observed in the $La_{1-x}Ca_xMnO_3$ with lower doping levels [57], thus suggesting an magnetic phase inhomogeneity near $T_C$ [58]. Such phase inhomogeneity scenario should be invoked to the understanding of unusual magnon behavior and transport properties in manganites including CMR effects. Remarkably, the central diffusive component maximizes its intensity very close to $T_C$, in a manner similar to the evolution of resistivity as well as the lattice polarons [26, 55, 59]. Figure 20 shows that the $T$-dependence of central component at $Q = (1.03, 0, 0)$ as well as the lattice polaron satellite peak at $Q = (3.75, 0.25, 0)$ through the FM phase transition is virtually identical to the evolution of resistivity in $La_{0.7}Ca_{0.3}MnO_3$, indicating that they all have a common origin which could be related to phase separation. Undoubtedly, such phase inhomogeneities should drastically affect the magnon lifetime, reflected by the unusual damping and the evolution of linewidth of magnons as $T \to T_C$.

## 9. Discussion on possible explanations

The main issues for the understanding of magnons and associated spin daynamics in FM half-metallic manganites include: *the unusual D-$T_C$ relation, the anisotropic zone-boundary softening and its dependence on doping, A-site disorder, as well as the anomalous zone-boundary broadening/damping*. There are quite a few theoretical studies attempting to explain these unusual magnon behaviors. Most of them have focused on the magnon softening and broadening. Although the DE interaction is still the basic ingredient for the understanding of spin dynamics in the FM metallic manganites, at least three major classes of theoretical approaches beyond the canonical DE interaction haven been proposed. The first class is based on the DE interaction and under the ferromagnetic Kondo lattice model, considering the effects of finite Hund's coupling [60], quantum and thermal corrections [12], on-site Coulomb repulsion [11], three-body correlation [61], conducting electron band ($e_g$) filling dependence of the DE and superexchange interactions [42, 62], and non-Stoner Continuum in the DE model [63]. The second class emphasizes the effect due to the quantum fluctuations of different $e_g$-orbitals [64, 65, 28]. The third ones goes beyond electronic origin by taking into account magnon-phonon coupling [46, 47, 66] and other lattice-related effects such as *A*-site disorder in the spin excitations [35].

As we already described in Sec. 2, the first attempt using an effective Kondo lattice model [9, 10] under the limit of $t/J_H \rightarrow 0$, which is equivalent to the nearest-neighbor FM Heisenberg model, gives a reasonable explanation of the magnon dispersion in $La_{0.7}Pb_{0.3}MnO_3$. This simple model fails to describe the anomalous magnon behaviors in relatively lower $T_C$-manganites, including $D$-$T_C$ relation, zone boundary softening and broadening. However, Shannon *et al.* [12] argued that the DE-ferromagnet, the ferromagnetic interaction between core spins mediated by conduction electrons (Kondo-type coupling), is generally not equivalent to the FM Heisenberg model. Specifically, the simple Heisenberg model is valid only when magnons are noninteracting quasiparticles and the spin waves are exact eigenstates of the Heisenberg Hamiltonian. Both quantum and thermal corrections to the magnetic properties of a DE model differ from any effective Heisenberg model because its spin excitations interact only indirectly, through the exchange of charge fluctuations. These new corrections do explain a doping (*x*)-dependent zone boundary magnon softening as compared to that in a Heisenberg ferromagnet. Yet, the corrections also predicts a relative magnon hardening in [1,1,1] direction at $T = 0$ K, which apparently is inconsistent with experimental observations [27].

Golosov [11] constructed a 1/*S* spin wave expansion [9] for DE ferromagnets with a sufficiently large Hund's rule coupling in a FM-metallic ground taking into consideration of the on-site coulomb repulsion [67]. He found that magnon-electron scattering, which gives rise to the subleading terms in the 1/*S* expansion, could provide corrections to the magnon dispersion and damping as well as their momentum dependences which are mediated by the Fermi surface geometry. In particular, it was found that the magnon linewidth $\Gamma(\vec{q})$ in the long-wavelength limit is proportional to $q^6$ in a three dimensional system such as $R_{1-x}A_xMnO_3$. This result is in agreement with the calculation by Shannon *et* *al.* [12] but different from the $q^4$-dependence based on magnon-magnon scattering [45]. Furthermore, it was predicted that an anomaly of $\Gamma(\vec{q})$ reflected by either a jump or a logarithmic divergence should occur at $q = k_F$. This seems in agreement with the observation of magnon linewidth anomaly near $\xi \sim 0.3$ [27]. However, the prediction of a strong doping dependence of *D* seems inconsistent with the experimental result (see Fig. 13 a) [21]. Also, the lack of detailed knowledge of the electronic band structure, especially the lack of reliable experimental data, prevents from a quantitative comparison between theoretical and experimental results.

Another possible source to renormalize the self-energy of magnon quasiparticles from the simple DE model is the excitations of "non-Stoner" continuum states. As we mentioned in Sec. 7, a prevalent view is that in FM half-metallic manganites the Stoner continuum (in the single-spin-flip channel) lies completely above the magnon energies due to larger Hund's rule coupling, thus keeping magnons from decaying into the continuum. However, exact diagonalization studies of the DE model by Kaplan *et al.* indicate that this prevalent picture is incorrect. Instead, there is a continuum of states closed to the magnon energies even for $J_H \to \infty$, and they probably overlap with magnons, providing magnon decay channels thus giving a contribution to the observed finite lifetime of magnons in inelastic neutron scattering. This also confirms the results obtained by Golosov [11]. Yet, *no* quantitative comparison between the theoretical calculations and experimental results is available yet.

Using the ferromagnetic Kondo lattice model but taking into account strong on-site correlations between $e_g$ electrons and AF exchange couplings among $t_{2g}$ spins, Mancini *et al.* [62] discovered that the competing FM, DE, and AF super-exchange interactions lead to a strong deviation of magnon dispersion close to the zone boundary from the spectrum obtained by the isotropic Heisenberg model. However, in *contrast* with the experimental observation [27], the calculational results indicate that magnons in the [1,1,1] direction should have the largest softening from that predicted by the nearest-neighbor Heisenberg model [see Eq. (7)] as compared with these in the [1,0,0] and [1,1,0] directions.

Both Khaliullin *et al.* [64] and Maezono *et al.* [65] realized the importance of the $e_g$-orbital degrees of freedom [68, 69] in the spin dynamics in FM metallic manganites. The strength of the FM interaction at a given bond strongly depends on the orbital character of $e_g$ electrons. Thus quantum fluctuations of $e_g$-orbitals are shown to strongly modulate the magnetic exchange bonds, thereby causing a renormalization of magnon dispersion. In particular, the short-wavelength magnons are affected because they are most sensitive to these local orbital fluctuations. This causes the unusual zone boundary magnon softening.

Khaliullin *et al.* [64] found that the presence of J-T phonons further enhance the orbital fluctuation effect. They considered J-T coupling of orbitals to lattice, which imposes low phonon frequencies onto orbital fluctuations, thus providing the phononic contribution to the magnon self-energy. With [64] or without [65] the involvement of the J-T phonons, the quantum fluctuations of the planar orbitals (i.e., $d_{x^2-y^2}$, $d_{y^2-z^2}$, and $d_{z^2-x^2}$) are dominant in the DE interaction, thus leading to the large anisotropy of spin dynamics. As a result, the magnon dispersion is largely renormalized in the [1,0,0] and [1,1,0] directions while very little affected in the [1,1,1] direction [64]. The magnons along the [1,1,1] direction are sensitive to all three spatial directions of exchange bonds, therefore, remain unaffected by the local symmetry breaking induced by low-dimensional orbital correlations [64, 70]. This gives an explicit explanation for the anisotropic magnon softening.

Considering the doping dependence of different orbital correlations which mediate the exchange couplings, Endoh *et al.* [28] argued that the anomalous magnon dispersion can be described by the phenomenological Heisenberg model with extended exchange coupling constants (*Js*). The theoretical results based on the local density approximation + Hubbard U calculations identified the contributions on exchange coupling constants from different type of orbital states. For $Sm_{0.55}Sr_{0.45}MnO_3$ with a relatively higher doping level of $x$ = 0.45, $J_4$ is enhanced considerably by the $d_{3z^2-r^2}$ orbital state [28] rather than the $d_{x^2-y^2}$ one for these with lower doping level of $x$ = 0.3 [64]. Nevertheless, the prediction on the doping dependence of $J_4/J_1$ is not quantitatively consistent with experimental results [21] as we have mentioned in Sec. 6.

Solovyev *et al.* [42] argued that the zone boundary magnon softening and the increase of $D$ with doping ($x$) have a purely magnetic spin origin. The observed magnon softening, demonstrating the importance of the long-range FM coupling, is a natural consequence of the $e_g$- band filling in the half-metallic regime, implying that the canonical DE limit ($t/J_H \to 0$) is not appropriate and that neither the lattice deformation nor the orbital ordering is required for the softening. Based upon the minimal tight-binding calculation including the consideration of the super-exchange interactions between localized spins, they found that the ratio of the longer-range coupling ($J_s$) to $J_1$ including $J_4/J_1$ depends on doping ($x$). However, the calculated doping-dependence of the zone boundary softening along the [1,0,0] direction seems to be much less than experimental results (see Fig. 13c). While it is yet to be confirmed about the agreement between the experimental and theoretical results of the magnon dispersion along the [1,1,0] and [1,1,1] directions.

Magnon-phonon coupling is another candidate beyond the electronic origin to renormalize the self-energy of magnons. In many low $T_C$ manganites, in which unusual magnon behavior has been observed, magneto-elastic and dynamic J-T effects [71] are believed to be crucial to transport and magnetic properties. In addition and as we described above, the observed unusual magnon behaviors, especially the magnon softening and broadening, have some relations to the J-T active phonons [23, 27, 44, 49]. Therefore, it is 'natural' to suggest that magnons are coupled to phonons.

Furukawa [46] has given a qualitative argument on the effects of magnon-phonon coupling for the case where their dispersions cross each other. When the interaction conserves the spin quantum number, the magnon linewidth becomes broad only when the magnon energy is higher than the magnon-phonon crossing energy. This seems to explain the observed magnon anomalous broadening close to zone boundary. A more realistic calculation on both phonon and magnon damping was given by Woods [47]. The calculation is based upon a model Hamiltonian in which the magnon part is taken care of by the Heisenberg Hamiltonian and the magnon-phonon coupling is reproduced by the scattering of a magnon with an emission or absorption of a phonon. In this case the coupling manifests itself through the distortion of lattice. By calculating the self-energy of magnons, the magnon softening and broadening have been reproduced. Due to the fact that the magnon damping is proportional to the boson population, this model explains in a natural way the enhanced damping with increasing temperature. However, the magnon-phonon coupling picture has difficulty in explaining the doping-dependence of $J_4/J_1$ (see Fig 13) and insensitivity of $J_4/J_1$ on average $A$-site ionic radius ($\bar{r}$) (see Fig 12). As the phonon frequency does not vary drastically with increasing doping (for $0.25 < x < 0.45$), a large change of $J_4/J_1$ with increasing doping would not be expected. On the other hand, if phonons play a crucial role, there would be a correlation between $T_C$ or $D$ and electron-phonon coupling strength ($g$) [71 - 74, 66]. While increasing $\bar{r}$ leads to rapid changes in $T_C$, it is hard to image the insensitivity of $D$ or $J_4/J_1$ on $T_C$ under the magnon-phonon scenario.

Finally, understanding the effect of quenched disorder on spin dynamics in FM metallic manganites is still an issue. It is known that the randomness of $A$-site substitution has drastic effect on $T_C$ as well as transport properties [39], suggesting that the disorder in the mixture of different size ion scatters the itinerant electrons and suppresses their kinetics. Motome *et al.* [35] have studied the spin excitation spectrum in the DE model with the presence of disorder. They found that the disorder causes anomalies in magnon spectrum including broadening, branching, anticrossing with gap opening. The $2k_F$ Friedel oscillation of spin and charge density in the fully polarized FM state caused by disorder which scatters the magnons and results in anticrossing in their dispersion etc. is believed the origin of these anomalies. According to their study, the increase of on-site disorder should enhance the magnon softening and damping. However, except the observed zone boundary broadening which may be used to compare with the theoretical results, the insensitivity of $D$ or $J_4/J_1$ on $\sigma^2$ seems to rule out the possibility of on-site disorder-induced zone boundary softening (see Fig. 12a).

## 10. Summary

In this review we have described recent studies of the magnon behaviors in FM metallic manganites. We focused only on metallic perovskite manganites $R_{1-x}A_xMnO_3$ (e.g. R = La, Nd, Pr, A = Sr, Ca, Pb), although there are a lot of excellent studies of the spin dynamics in non-metallic phase of these three-dimensional materials as well as the spin dynamics of layered manganite compounds. In spite of a great deal of experimental and theoretical effort, a clear picture of the spin dynamics including its doping and temperature dependence is yet to emerge. Theoretically, several mechanisms proposed so far have certain degrees of success accounting for anomalous magnon behaviors deviating from the simple canonical Heisenberg model. However, none of them can satisfactorily explain all of the observed results. Experimentally, considerably more work is needed to fully characterize the momentum-, temperature- and doping-dependence of magnons. In particular, few important experimental measurements are essential to further test these theoretical mechanisms:

1) The possible zone boundary magnon softening in high-$T_C$ manganites as compared with the simple Heisenberg model with nearest-neighbor interaction, in order to confirm that whether the zone boundary softening is a universal phenomenon in all FM metallic manganites.

2) The full characterization of magnon dispersion in the [1,1,1] direction, especially for the low-$T_C$ manganites. This requires high energy neutrons.

3) The systematic measurements on the doping- and temperature-dependence of magnon linewidth $\Gamma(\vec{q})$.

4) The Isotope effects on magnon dispersion and damping to further testing the magnon-phonon coupling.

5) The scaling behavior of the spin wave stiffness $D(T)$, especially the magnon non-collapsing issue when $T \to T_C$.

6) The systematic measurement on the electronic band structure including the Fermi surfaces which is crucial to understand the correlation effects on spin dynamics. Though progress in the study of the Fermi surface topology with high-resolution angle-resolved photoelectron spectroscopy (ARPES) has been made in the layered manganites such as $La_{1.2}Sr_{1.8}Mn_2O_7$ [75, 76], it is difficult to determine the electronic structure in these three dimensional manganites because they cannot be cleaved to obtain a reasonably good surface. Surface effects [77, 78] due to the surface lattice relaxation, segregation, and/or imperfection would affect the measured electronic structure by using surface sensitive techniques like ARPES.

Polarized inelastic neutron scattering will be extremely helpful for distinguishing the magnon excitations from others and extracting the truly intrinsic magnon bandwidth and linewidth, especially near the zone boundary where magnons merge with phonons. Meanwhile, the effect of lattice distortion, quenched disorder, and even phase separation, should be important, especially in the temperature closed to $T_C$, thus deserving further investigation. Due to the nature of the close and complex coupling between charge, lattice, orbital and spin degrees of freedom in this class of materials, how to tailor these different interactions for revealing their effects on spin dynamics should be the main challenge.

**Acknowledgments**

This research was sponsored By the U.S Department of Energy (DOE), Basic Energy Sciences, Division of Materials Sciences and Engineering and conducted at Oak Ridge National Laboratory (ORNL).  JZ was also supported in part by the U.S. National Science Foundation (NSF) with Grant No. DMR-0346826 and PD was supported in part by NSF with Grant No. DMR-0453804. ORNL is supported by the U.S. DOE Grant No. De-AC05-00OR22725 through UT/Battelle LLC.

Table 1 A summary of the fit results of magnon dispersion data to the Heisenberg Hamiltonian with nearest-neighbor ($J_1$) and fourth-nearest-neighbor ($J_4$) exchange coupling. The Curie temperature ($T_C$), A-site disorder ($\sigma^2$), spin-wave stiffness (D), $J_1$, $J_4$, and their ratio ($J_4/J_1$) are listed. References where the data are taken from are also shown in column for D.

| Samples | $T_C$ (K) | $\sigma^2$ (x10$^{-3}$) | D [ref.] (meV Å$^2$) | $2SJ_1$ (meV) | $2SJ_4$ (meV) | $J_4/J_1$ (%) |
|---|---|---|---|---|---|---|
| La$_{0.7}$Sr$_{0.3}$MnO$_3$ | 378 | 1.8556 | 188 [18]<br>176 [19] | 7.63 | 1.66 | 22 ± 1.4 |
| La$_{0.7}$Pb$_{0.3}$MnO$_3$ | 355 | 3.7708 | 134 [17] | 8.79 | --- | --- |
| La$_{0.7}$Ba$_{0.3}$MnO$_3$ | 350 | 1.3548 | 152 [32] | 7.30 | 1.42 | 20 ± 3.2 |
| La$_{0.68}$Ba$_{0.32}$MnO$_3$ | 336 | 1.4038 | --- [31] | 7.03 | 1.59 | 23 ± 2.1 |
| La$_{0.8}$Sr$_{0.2}$MnO$_3$ | 312 | 1.4138 | 167 [19] | --- | --- | --- |
| Pr$_{0.63}$Sr$_{0.37}$MnO$_3$ | 301 | 4.0002 | 165 [24] | 5.16 | 2.08 | 40 ± 4.4 |
| La$_{0.67}$Ca$_{0.33}$MnO$_3$ | 250 | 0.2865 | 170 [26] | | | |
| La$_{0.7}$Ca$_{0.3}$MnO$_3$ | 238 | 0.2722 | 165 [27] | 6.63 | 0.76 | 11 ± 1.4 |
| Nd$_{0.7}$Sr$_{0.3}$MnO$_3$ | 198 | 4.5379 | 165 [25] | --- | --- | --- |
| La$_{0.75}$Ca$_{0.25}$MnO$_3$ | 191 | 0.2430 | 152 [21] | 7.90 | 0.36 | 5 ± 0.5 |
| Sm$_{0.55}$Sr$_{0.45}$MnO$_3$ | 135 | 7.8418 | 140 [28] | 2.91 | 1.74 | 60 ± 8.9 |
| Pr$_{0.7}$Ca$_{0.3}$MnO$_3$ | * | 2.1E-04 | 145 [21] | 5.98 | 0.61 | 10 ± 2.6 |
| Pr$_{0.55}$(Ca$_{0.85}$Sr$_{0.15}$)$_{0.45}$MnO$_3$ | * | 2.0973 | 152 [21] | 3.24 | 1.79 | 55 ± 8.1 |

* The ground state of these materials is AF insulating state but can be transformed into FM state by field cooling.

**Figure Captions:**

Fig. 1 Ball model of the crystal structure of typical FM-metallic $R_{1-x}A_xMnO_3$ manganites with pseudo-cubic perovskite (orthorhombic) symmetry. Red balls represent cation elements, blue ones are oxygen ions while Mn ions are in the center of green cages.

Fig. 2 Pseudo-cubic crystal structure of $R_{1-x}A_xMnO_3$ with different Mn neighbors of magnetic exchange coupling indicated. Red balls represent cation elements, blue ones oxygen ions, and green ones Mn ions.

Fig. 3 Magnon dispersion of $La_{0.7}Pb_{0.3}MnO_3$ along the three major cubic symmetry directions at 10K determined by inelastic neutron scattering. Solid lines is the fit to the Heisenberg model with nearest-neighbor coupling $2J_1S = 8.79 \pm 0.21$ meV. The figure is taken from Ref. [17].

Fig. 4 Sr-doping dependence of $T_C/T_N$ and spin-wave stiffness $D$ in $La_{1-x}Sr_xMnO_3$. The dashed lines are guiders to the eye. The figure follows the plot of Fig.1 in Ref. [20] but new data points are added.

Fig. 5 Magnon dispersion of $Pr_{0.63}Sr_{0.37}MnO_3$ along the [0,0,1], [1,1,0], and [1,1,1] directions (the zone boundary is at $\xi = 0.5$) [24]. The solid line is a fit to a nearest-neighbor Heisenberg Hamiltonian for $T = 10$ K and $\xi < 0.2$ while the dashed curve is a fit for all data including up to the fourth nearest neighbor couplings at $T = 10$ K. The dotted line is the corresponding fit for $T = 265$ K. Also shown in squares are the data for $La_{0.7}Pb_{0.3}MnO_3$ (see Fig. 2 and Ref [17]).

Fig. 6 $T$-dependence of the resistivity $\rho(T)$ of the single crystal $Nd_{0.7}Sr_{0.37}MnO_3$, $La_{0.7}Ca_{0.3}MnO_3$, and $Pr_{0.63}Ca_{0.37}MnO_3$ for the magnon measurements [27]. The large drop in $\rho(T)$ corresponds to the $T_C$ at 198 K, 238 K, and 301 k, respectively. The inset shows the normalized resistivity $\rho(T)/\rho(0)$. Note that all three compounds have the same $T$-dependence of $\rho(T)/\rho(0)$ in the FM metallic state below 100 K.

Fig. 7 Sequence of the constant-*q* scans at selected wave vectors for the magnon excitations in $La_{0.7}Ca_{0.3}MnO_3$ along the [0,0,1] direction at $T = 10$ K. The solid curves are the Gaussian fits to the data and the dash lines represent the linear backgrounds.

Fig. 8 Magnon dispersions (open symbols) of $Nd_{0.7}Sr_{0.37}MnO_3$, $La_{0.7}Ca_{0.3}MnO_3$, and $Pr_{0.63}Sr_{0.37}MnO_3$ at $T = 10$ K along both [0,0,1] and [1,1,0] directions [27]. Solid symbols show the dispersion of selected LO-phonon modes collected along the reciprocal-lattice directions as specified in the legend.

Fig. 9 Phase diagrams of $Pr_{0.7}Ca_{0.3}MnO_3$ [33] and $Pr_{0.55}(Ca_{0.85}Sr_{0.15})_{0.45}MnO_3$ [21] in *T-H* plane based on transport measurement. The neutron scattering measurements on magnon excitations were taken at the position marked by the (red) upper triangle and the (blue) square symbols.

Fig. 10 *q*-dependence magnon excitation spectra in (a) $La_{0.75}Ca_{0.25}MnO_3$, (b) $Pr_{0.7}Ca_{0.3}MnO_3$ under 5 T magnetic field, and (c) $Pr_{0.55}(Ca_{0.85}Sr_{0.15})_{0.45}MnO_3$ under 7 T magnetic field. The spectra at different *q*'s are incrementally shifted for clarity. The instrumental resolutions are shown in the horizontal bars and the shoulders around 15 meV in (c) are the phonon scattering. *E* versus $q^2$ is plotted in the insets to determine *D*.

Fig. 11 Summary of magnon dispersion curves along the [1,0,0] direction for (a) various $R_{0.7}A_{0.3}MnO_3$ manganites and (b) a series of $R_{1-x}A_xMnO_3$ as a function of *x*. The solids are the least-square fits using the Heisenberg model with $J_1$ and $J_4$. Data are obtained from Ref. [19, 21, 24, 25, 27, 28, 32].

Fig. 12 Disorder dependence of (a) the spin-wave stiffness *D* measured from low-*q* magnon excitations, (b) the nearest-neighbor exchange coupling $2SJ_1$ and (d) the ratio $J_4/J_1$; The average ionic radius dependence of (d) *D*, (e) $2SJ_1$, and (f) $J_4/J_1$. Dashed lines are guides to the eye (from [21]).

Fig. 13 Doping (*x*)-dependence of (a) *D*, (b) $2SJ_1$, and (c) $J_4/J_1$. Dashed lines are guides to the eye. The solid curve in (a) is the prediction of Ref [11]. The (green) solid line in (c) is the result calculation result from Ref [42] while (blue) dash-dotted and (red) dashed lines in (c) are these from Ref. [28]. The figure is taken from Ref [21].

Fig. 14 Linewidth of magnon excitation spectra along the [1,0,0] direction for $Nd_{0.7}Sr_{0.37}MnO_3$, $La_{0.7}Ca_{0.3}MnO_3$, and $Pr_{0.63}Sr_{0.37}MnO_3$ at $T = 10$ K. a significant

magnon linewidth broadening is seen after $\xi \geq 0.3$ (marked by arrows) when an optical phonon merges with the magnon. The figure is taken from Ref [27].

Fig. 15 (A) Dispersion of magnons (solid symbols) along the [1,0,0], [1,1,0], and [1,1,1] direction and two related optical phonon modes (open symbols) along the [1,0,0] ($\Omega_1$) and [1,1,0] ($\Omega_2$) direction of La$_{0.7}$Ca$_{0.3}$MnO$_3$ at $T$ = 10 K; (B) the linewidth of magnon excitation spectra as a function of magnon energy (bottom) of La$_{0.7}$Ca$_{0.3}$MnO$_3$ at $T$ = 10 K. The inset shows the magnon linewidth as a function of energy in the [1,1,1] direction. The solid lines are the guides to the eye.

Fig. 16 *T*-dependence of magnon dispersion along the [1,0,0] direction of La$_{0.7}$Ca$_{0.3}$MnO$_3$.

Fig. 17 *q*-dependence of Magnon linewidths of Pr$_{0.63}$Sr$_{0.37}$MnO$_7$ along the [0,0,1] direction at $T$ = 10 K and 265 K. The solid and dashed lines are guiders to the eye. The figure is taken from Ref. [22].

Fig. 18 (*T/T$_C$*)-dependence of spin-wave stiffness $D(T/T_C)$ in Nd$_{0.7}$Sr$_{0.37}$MnO$_3$ ($T_C$ = 198 K) La$_{0.7}$Ca$_{0.3}$MnO$_3$ ($T_C$ = 238 K), and Pr$_{0.63}$Sr$_{0.37}$MnO$_3$ ($T_C$ = 301 K). A discontinuity in $D$ as $T \to T_C$ is obvious for Nd$_{0.7}$Sr$_{0.37}$MnO$_3$ and La$_{0.7}$Ca$_{0.3}$MnO$_3$.

Fig. 19 (Top) *T*-dependence below $T_C$ = 250 K and (bottom) field-dependence at T = 240 K of magnetic inelastic neutron scattering spectra for La$_{2/3}$Ca$_{1/3}$MnO$_3$ polycrystal. The figure is taken from Ref. [52]. The central peak is due to quasi-elastic spin diffuse scattering while the two side peaks are due to the magnon excitations. Similar results have been obtained from La$_{0.7}$Ca$_{0.3}$MnO$_3$ single crystal [55]. The curves are fits to data.

Fig. 20 *T*-dependence of the intensity of the central diffusive scattering peak of La$_{0.7}$Ca$_{0.3}$MnO$_3$ single crystal, compared with that of the polaron peak at a wave vector of $\boldsymbol{Q}$ = (3.75, 0.25, 0), and with the sample resistivity. The data have been scaled so that the peak heights match. The similarity of the data suggests a common physical origin. The figure is taken from Ref. [55]. Similar results have obtained in Ref [59].

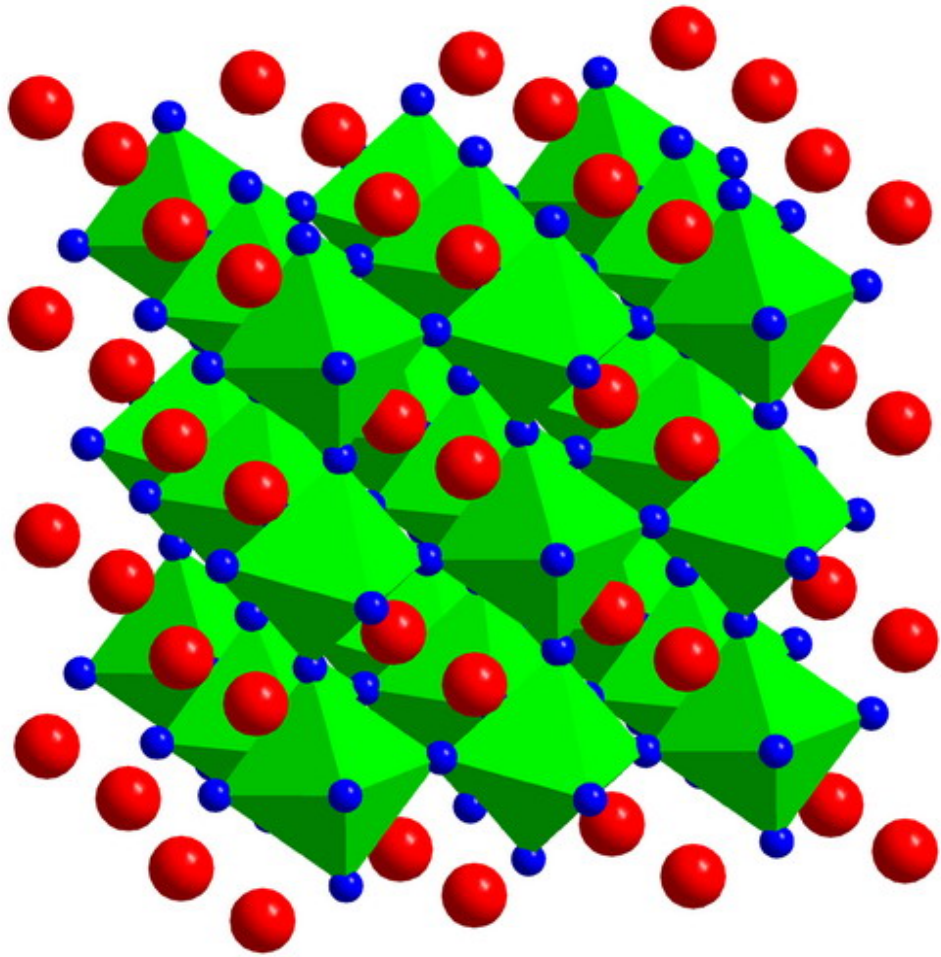

**Fig. 1**

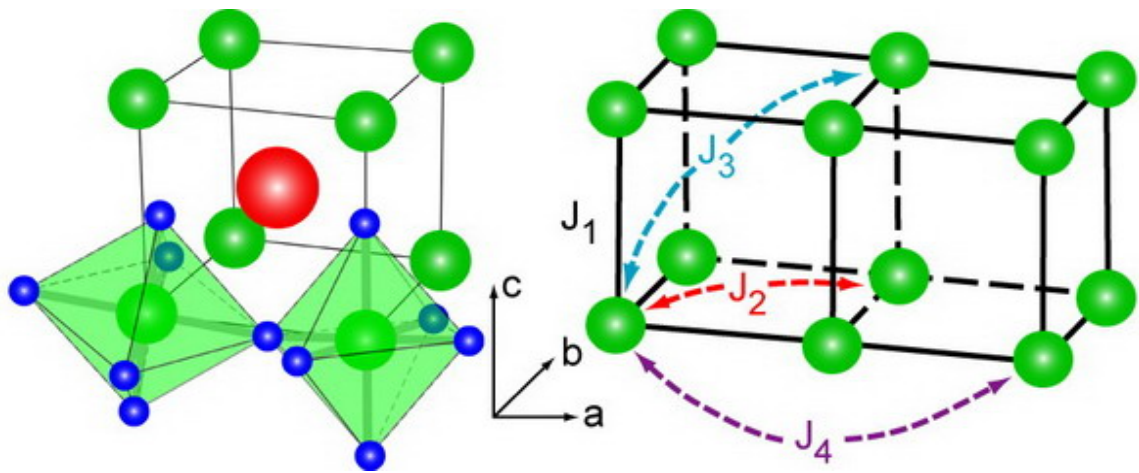

**Fig. 2**

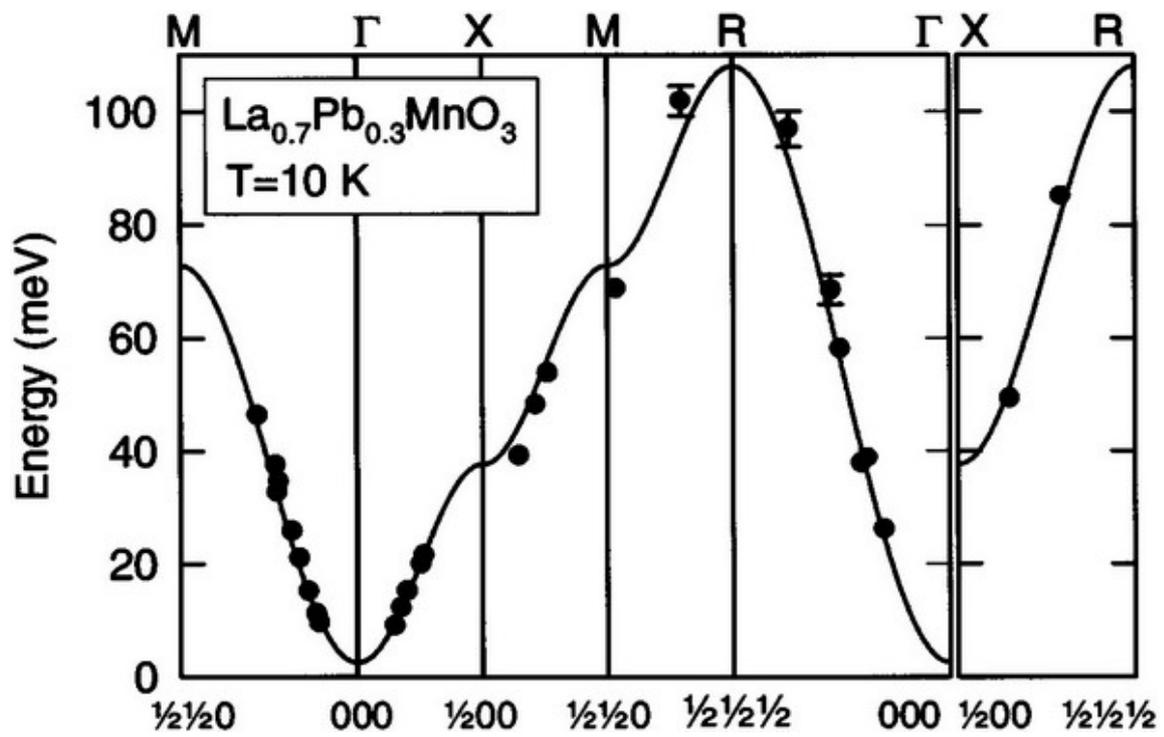

**Fig. 3**

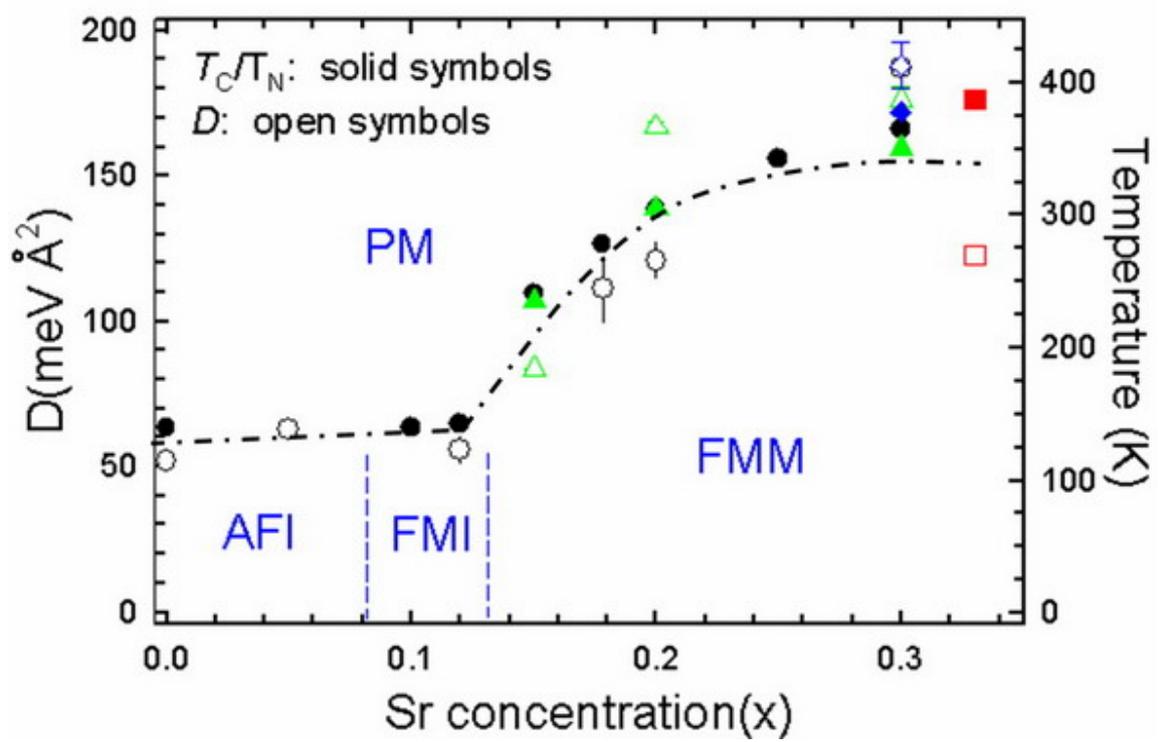

**Fig. 4**

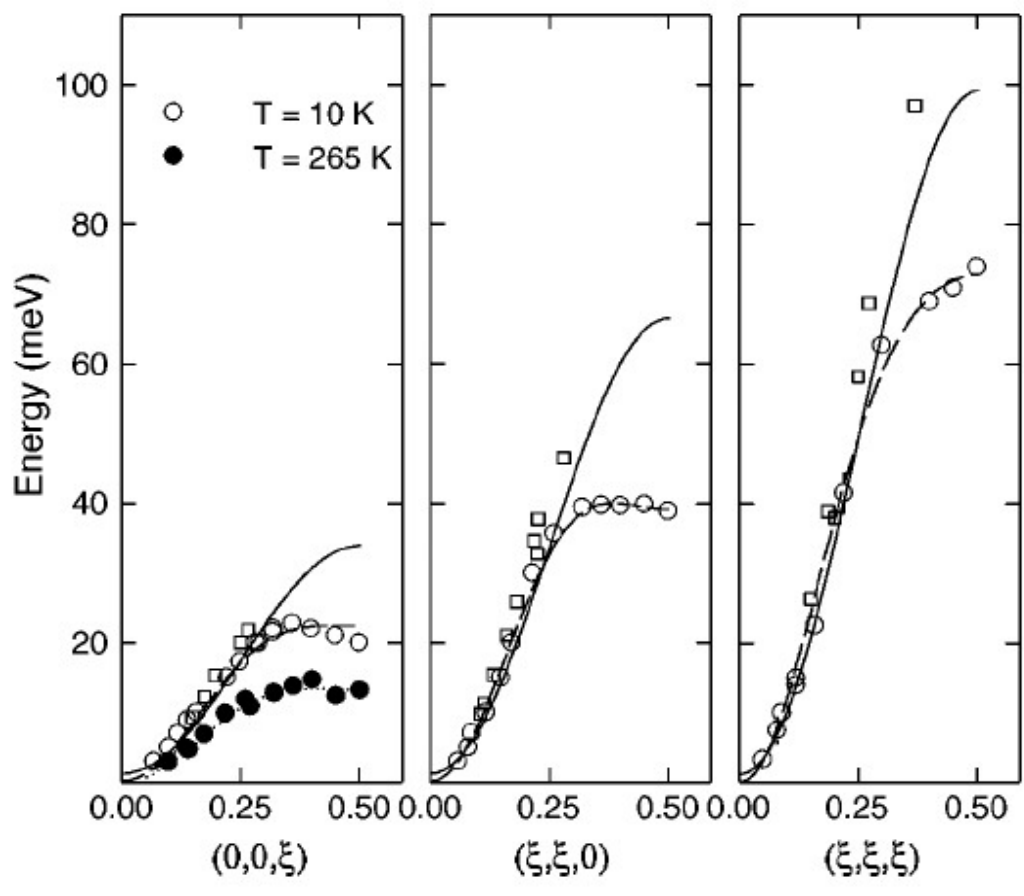

**Fig. 5**

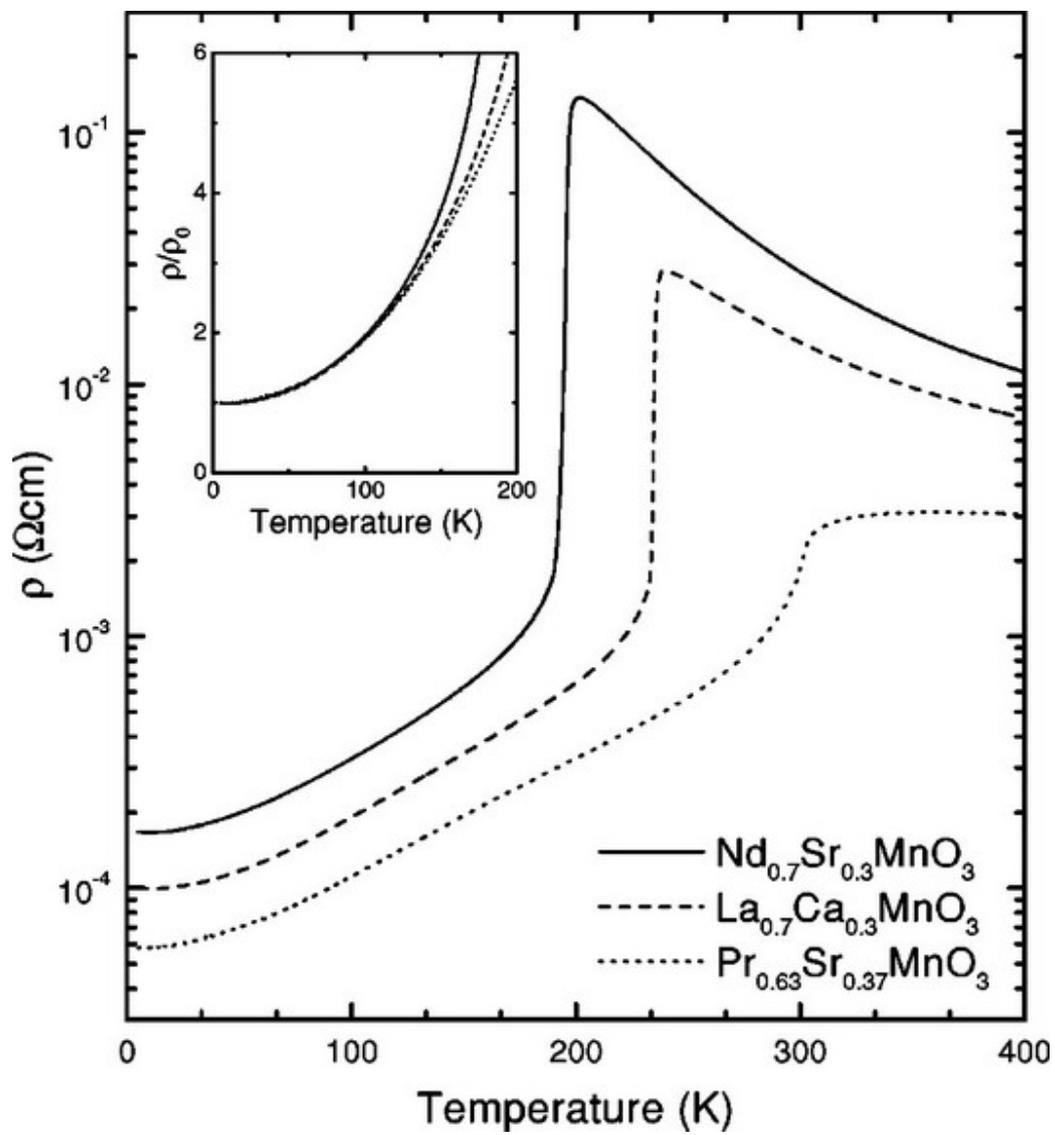

**Fig. 6**

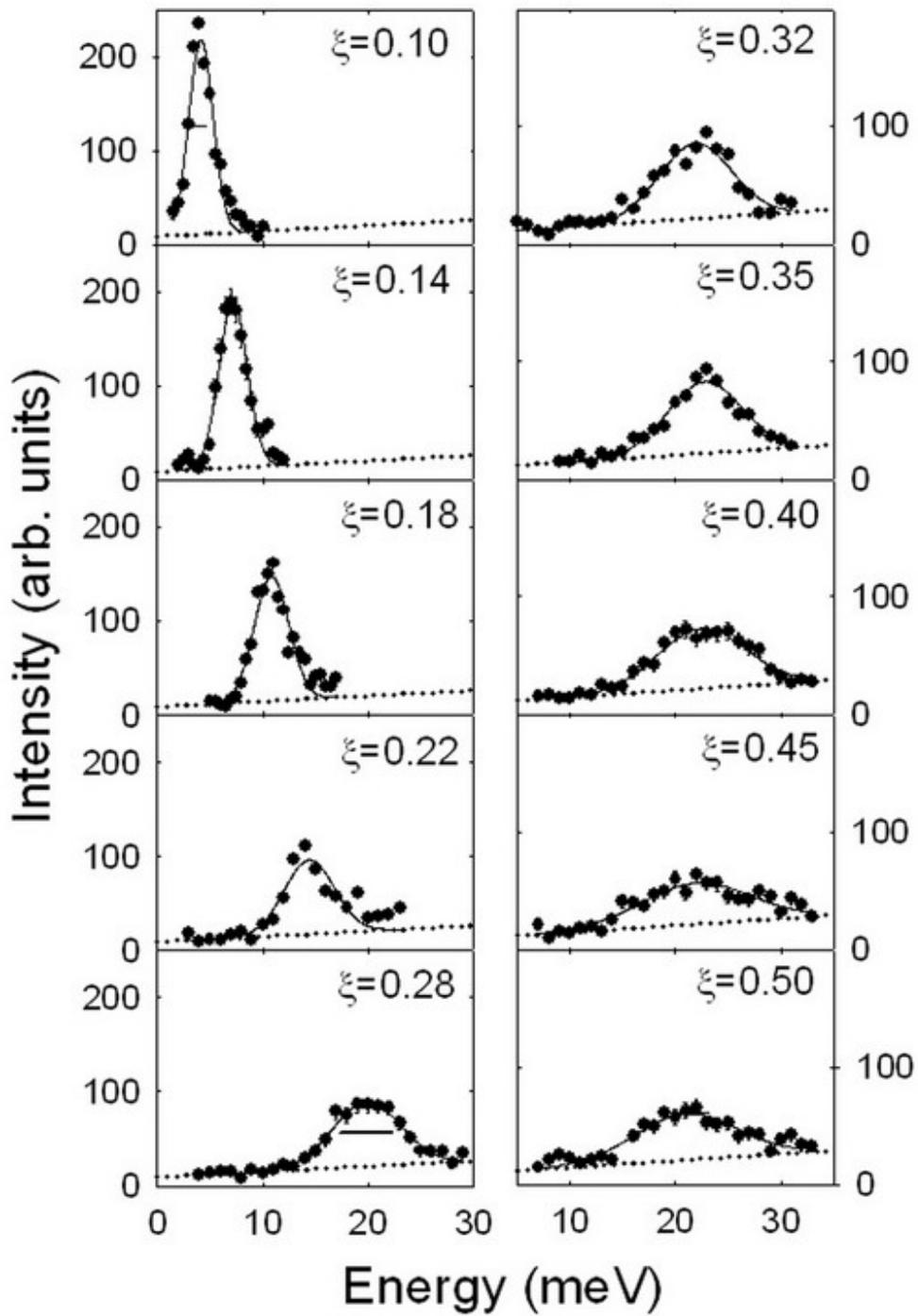

**Fig. 7**

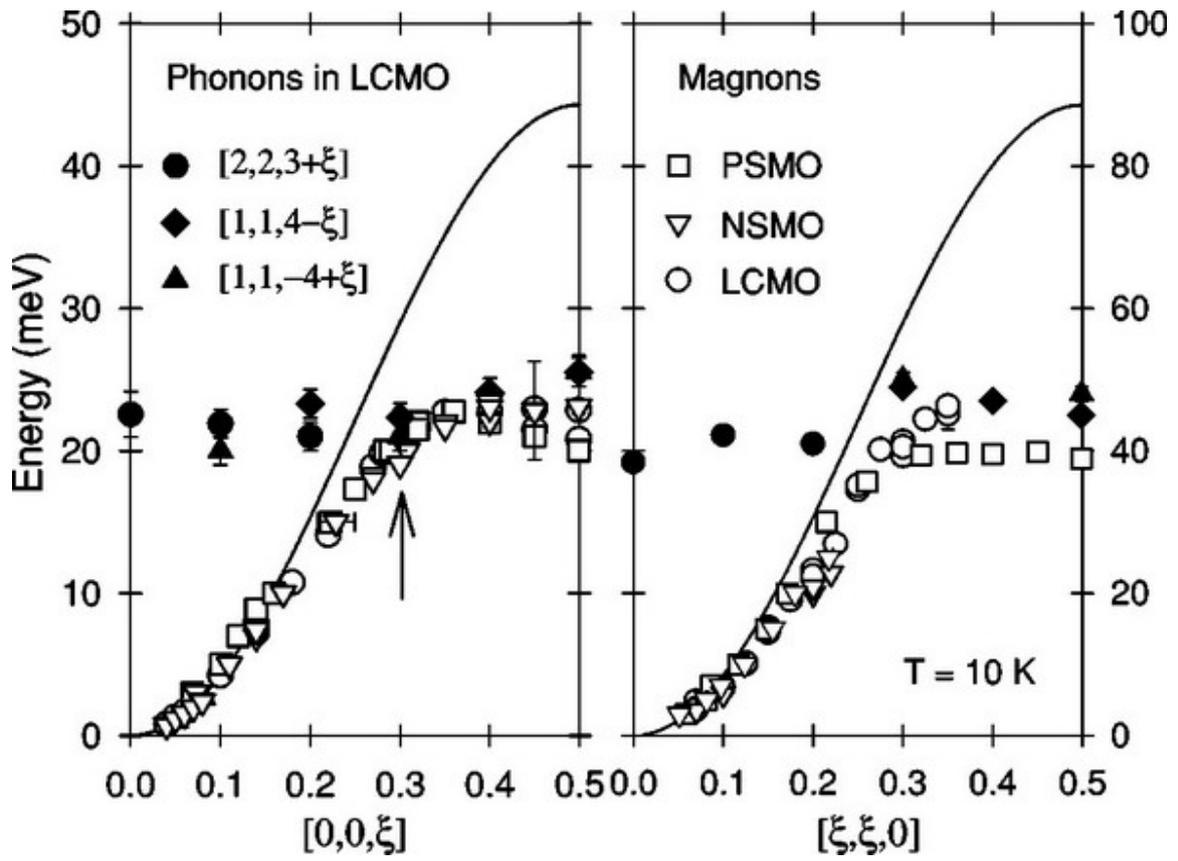

**Fig. 8**

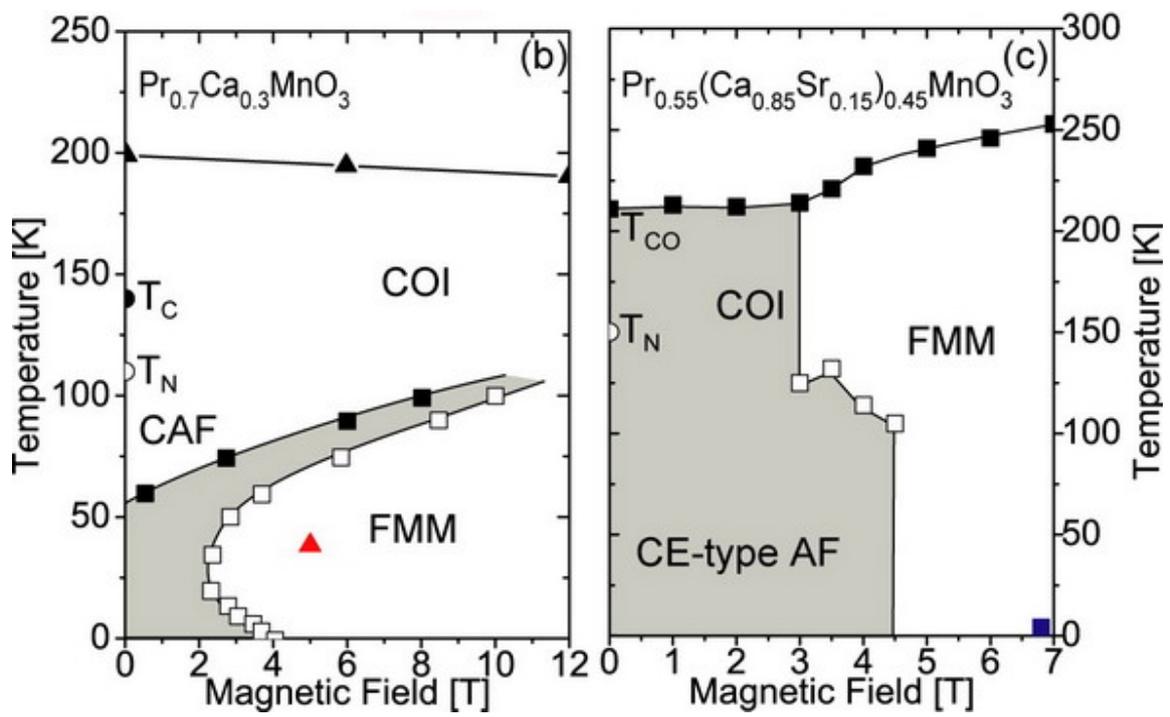

Fig. 9

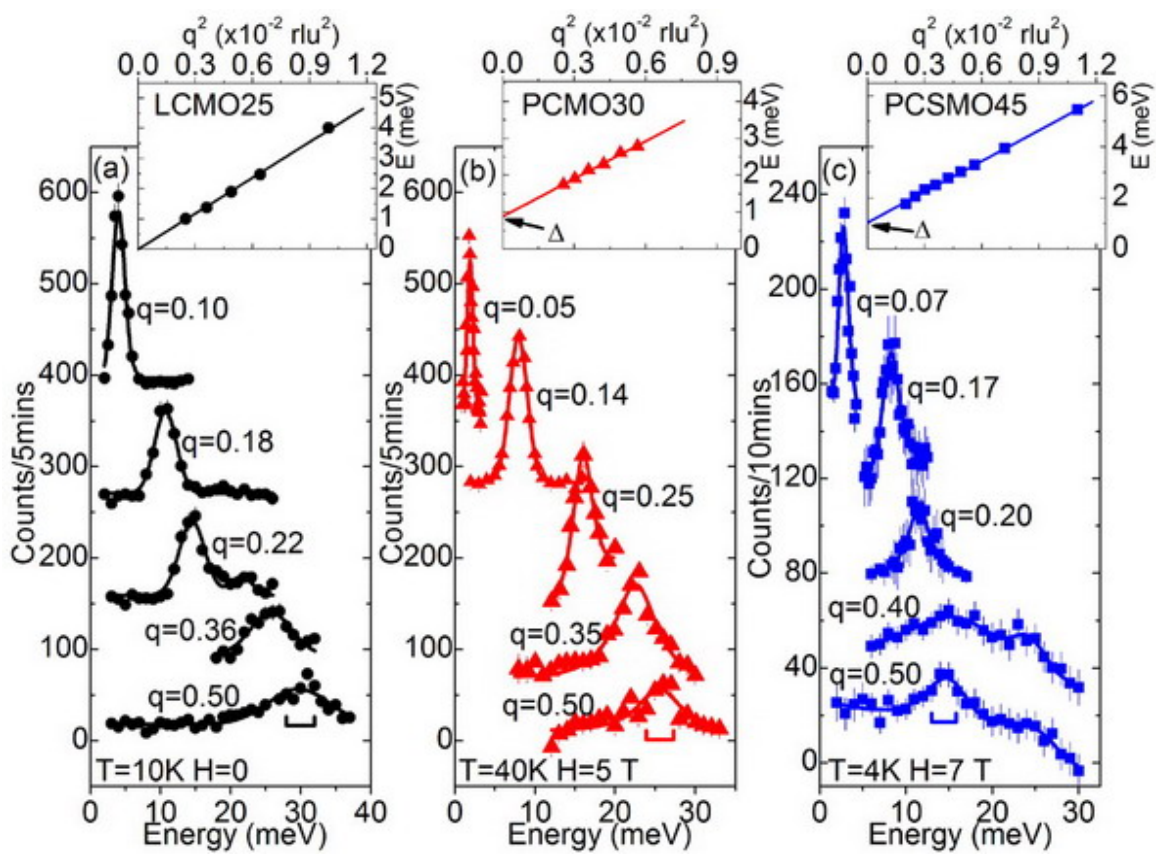

Fig. 10

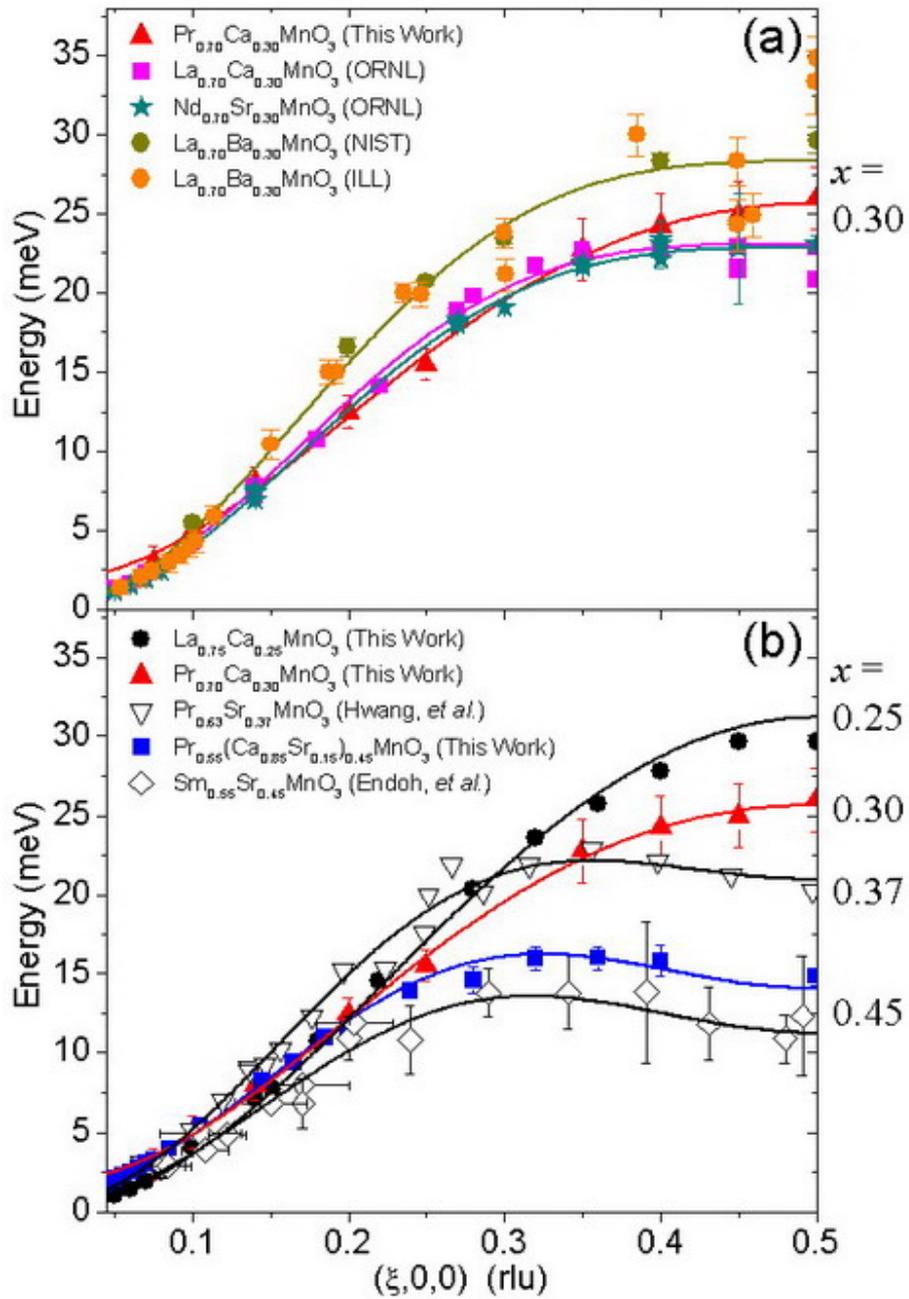

**Fig. 11**

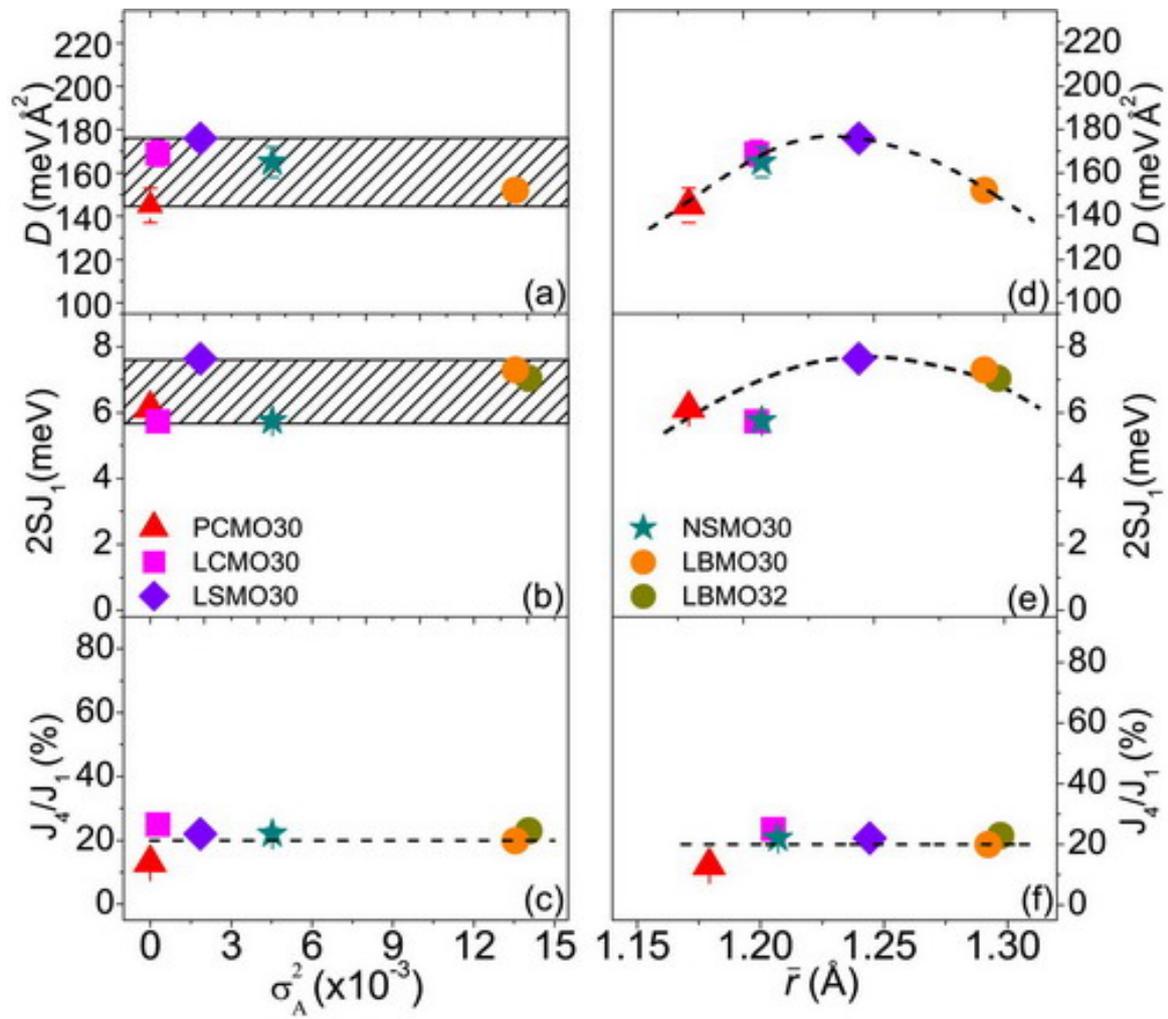

**Fig. 12**

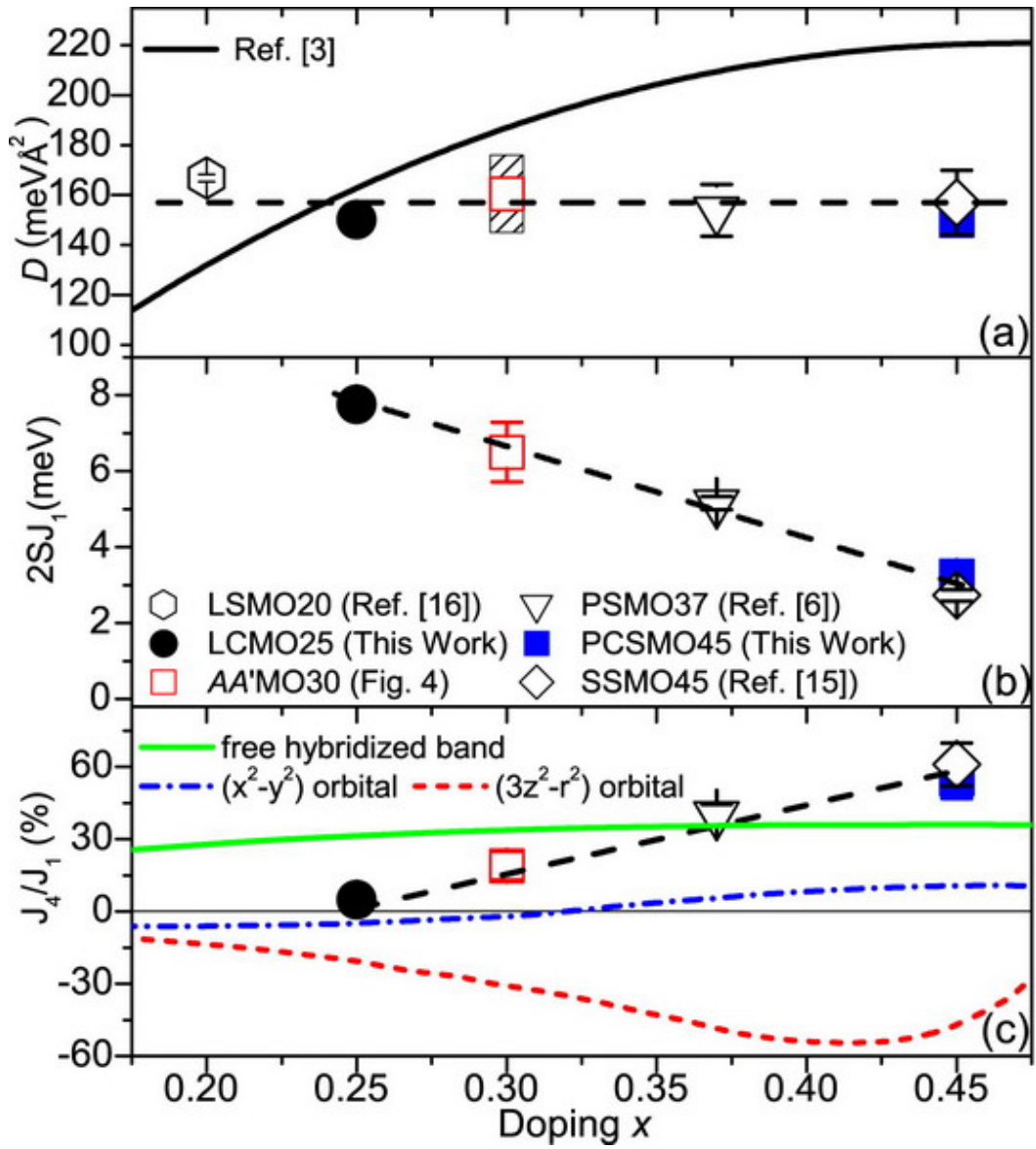

**Fig. 13**

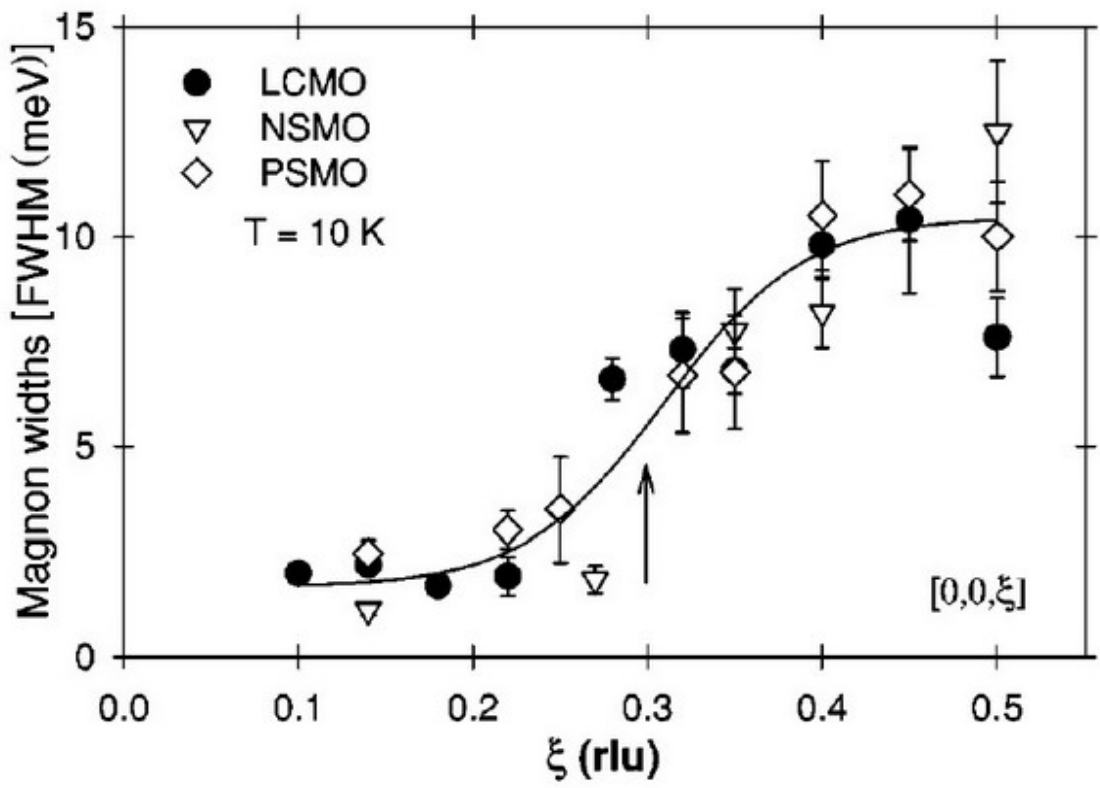

**Fig. 14**

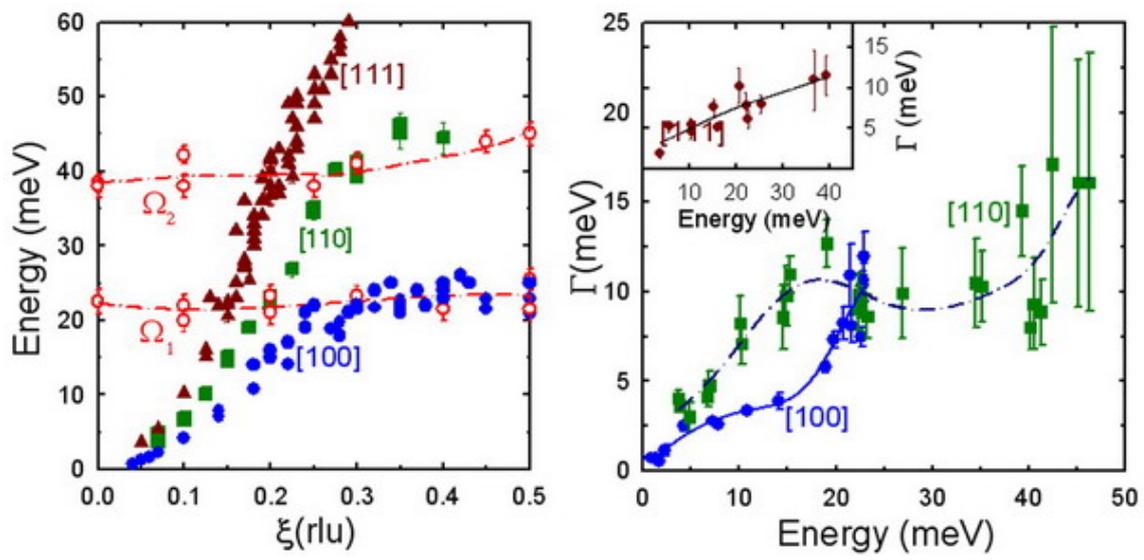

**Fig. 15**

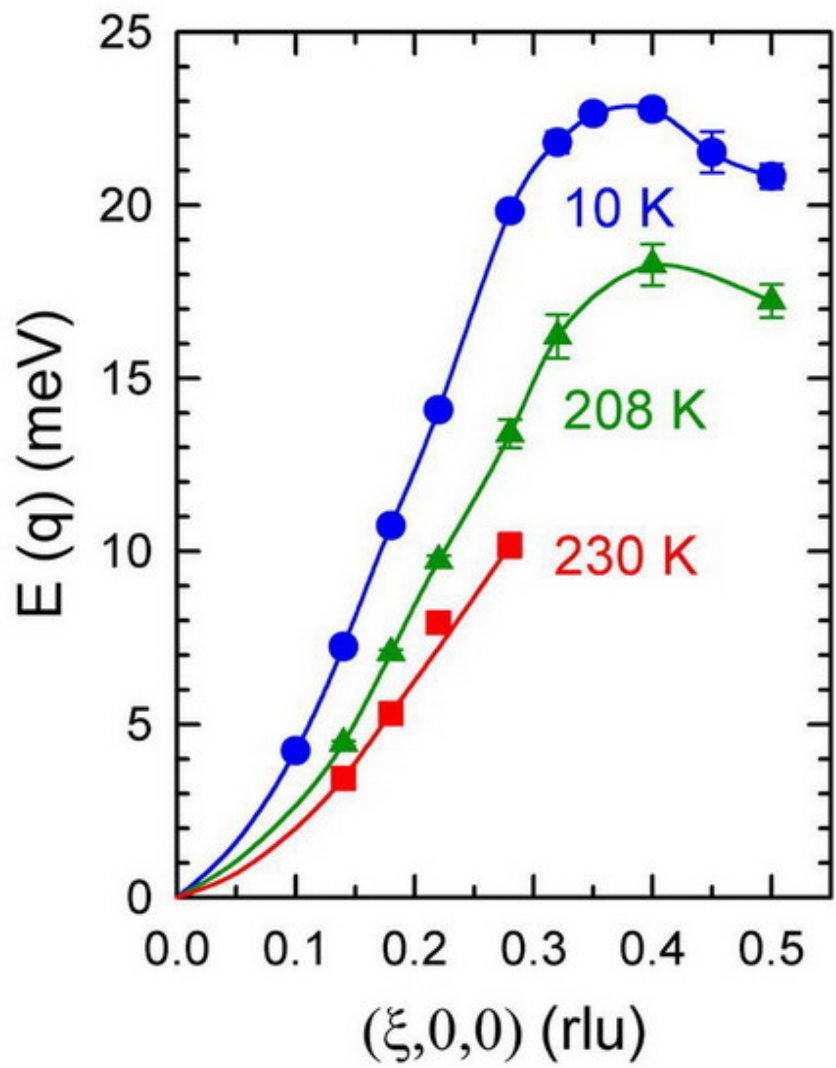

**Fig. 16**

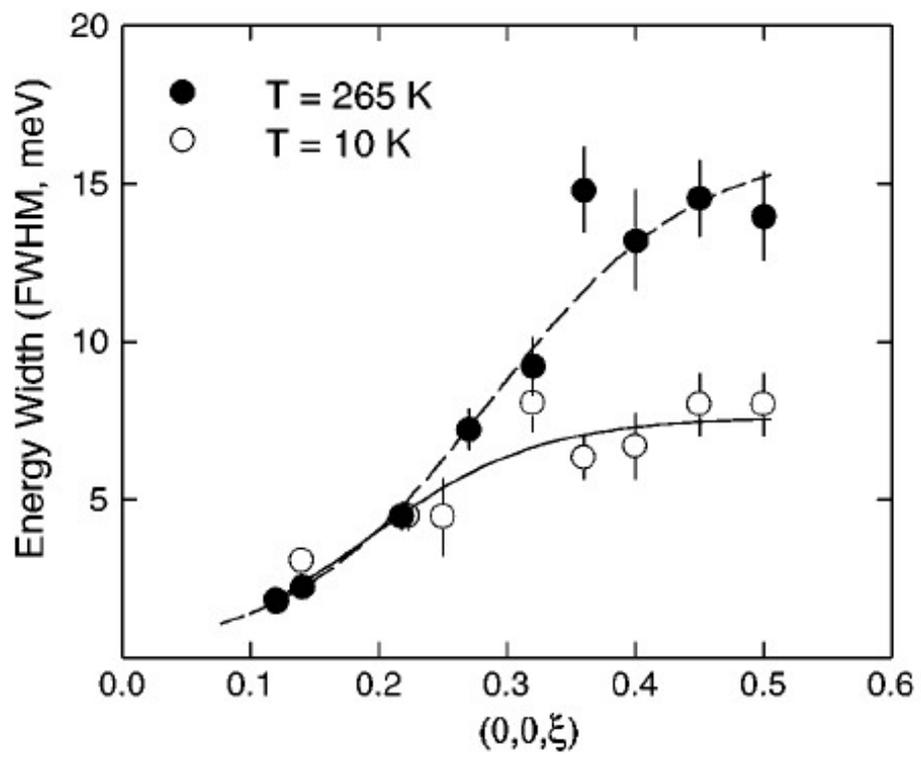

**Fig. 17**

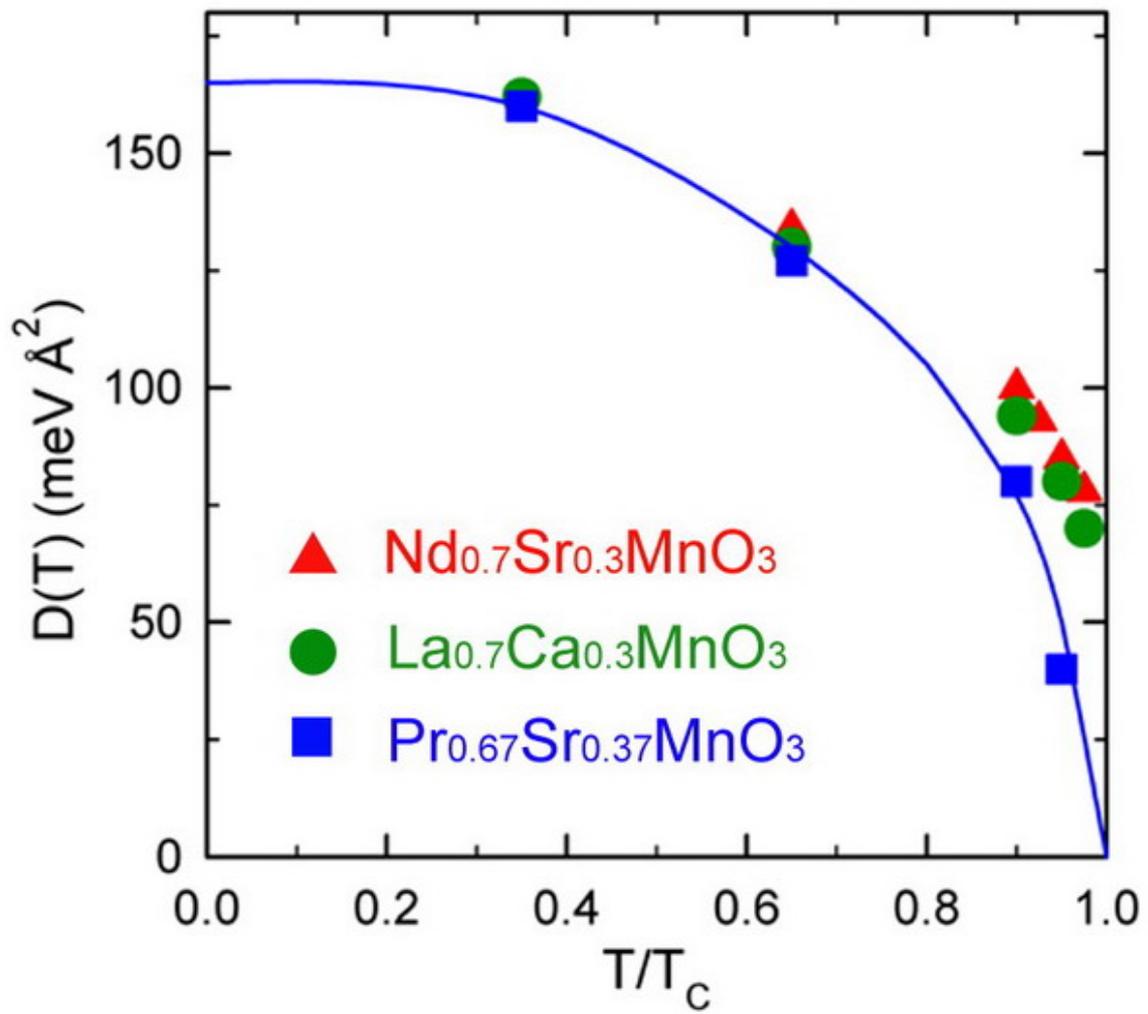

**Fig. 18**

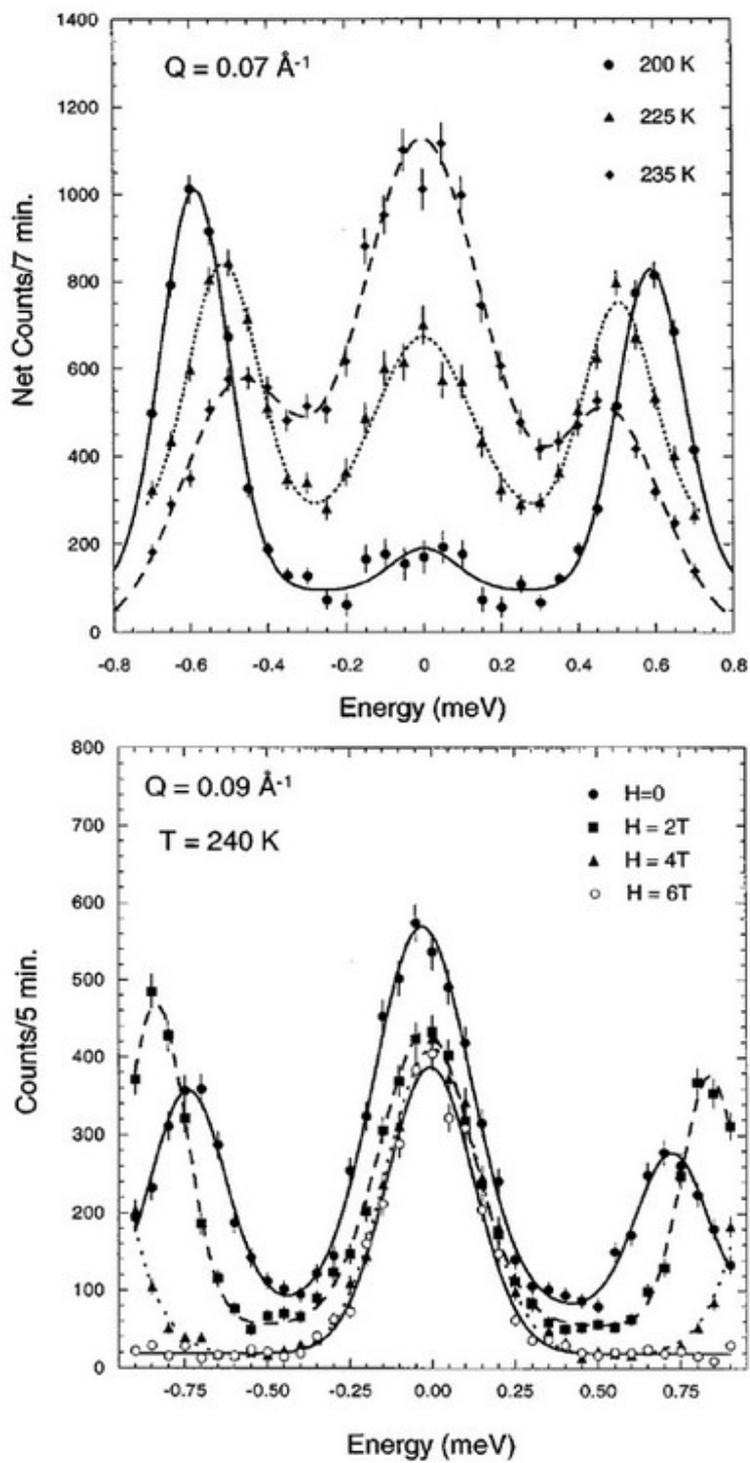

**Fig. 19**

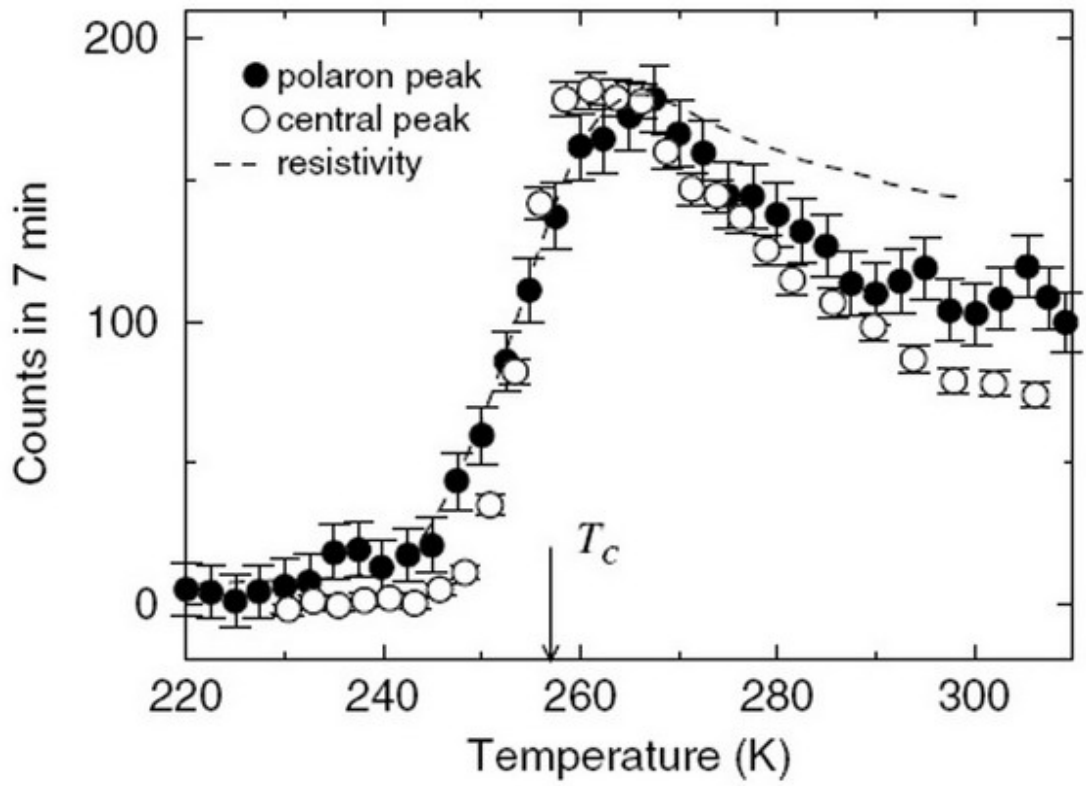

**Fig. 20**